\documentclass{aa}  

\usepackage{graphicx}
\usepackage{txfonts}
\usepackage[]{hyperref}

\usepackage{tablefootnote}
\usepackage{subfigure}
\usepackage[symbol]{footmisc}
\renewcommand{\thefootnote}{\arabic{footnote}}

\hypersetup{
    colorlinks=true,
    citecolor=blue,
    linkcolor=blue,
    filecolor=magenta,      
    urlcolor=cyan,
}

\usepackage{xcolor}

\defcitealias{bergamini19}{B19}
\defcitealias{bergamini22}{B23}
\defcitealias{granata22}{G22}

\usepackage{natbib}
\bibpunct{(}{)}{;}{a}{}{,} 
\begin{document} 

   \titlerunning{Exploring the low-mass regime of galaxy-scale SL}
   \authorrunning{G. Granata et al.}
   \title{Exploring the low-mass regime of galaxy-scale strong lensing: Insights into the mass structure of cluster galaxies}
   \author{G. Granata
          \inst{1,2},
          P. Bergamini
          \inst{1,3},
          C. Grillo
          \inst{1,2},
          M. Meneghetti
          \inst{3,4},
          A. Mercurio
          \inst{5,6},
          U. Me{\v{s}}tri{\'c}
          \inst{1,3},
          A. Ragagnin
          \inst{3,7},
          P. Rosati
          \inst{8,3},
          G.~B. Caminha
          \inst{9,10},
          L. Tortorelli
          \inst{11},
          \and
          E. Vanzella
          \inst{3}
          }

   \institute{Dipartimento di Fisica, Universit\`a degli Studi di Milano, Via Celoria 16, 20133 Milano, Italy\\
              \email{giovanni.granata@unimi.it}
              \and 
              INAF - IASF Milano, via Corti 12, 20133 Milano, Italy
              \and
              INAF - Osservatorio di Astrofisica e Scienza dello Spazio di Bologna, Via Piero Gobetti 93/3, 40129 Bologna, Italy
              \and
              INFN - Sezione di Bologna, Viale Berti Pichat 6/2, 40127 Bologna, Italy
              \and
              Dipartimento di Fisica “E.R. Caianiello”, Università Degli Studi di Salerno, Via Giovanni Paolo II, 84084 Fisciano (SA), Italy
              \and
              INAF - Osservatorio Astronomico di Capodimonte, Salita Moiariello 16, 80131 Napoli, Italy
              \and
              IFPU - Institute for Fundamental Physics of the Universe, Via Beirut 2, 34151 Trieste, Italy
              \and
              Dipartimento di Fisica e Scienze della Terra, Universit\`a degli Studi di Ferrara, Via Saragat 1, 44122 Ferrara, Italy
              \and
              Technische Universit\"at M\"unchen, Physik-Department, James-Franck Str. 1, 85748 Garching, Germany
              \and
              Max-Planck-Institut f\"ur Astrophysik, Karl-Schwarzschild-Str. 1, 85748 Garching, Germany
              \and
              Universit\"ats-Sternwarte, Fakult\"at f\"ur Physik, Ludwig-Maximilians-Universit\"at M\"unchen, Scheinerstr.1, 81679 München, Germany
            }

   \date{November 29, 2023}

  \abstract
   {Several recent studies have highlighted a discrepancy between the strong lensing (SL) properties of observed cluster galaxies and the predictions of \textrm{$\Lambda$} cold dark matter (CDM) cosmological hydrodynamical simulations. This discrepancy can be interpreted as the result of observed cluster members being more compact than their simulated counterparts.} 
   {In this work, we aim at a direct measurement of the compactness of a few selected galaxy-scale lenses in massive clusters, testing the accuracy of the scaling laws adopted to describe the members in SL models of galaxy clusters.}
   {We selected the multiply imaged sources MACS J0416.1$-$2403 ID14 ($z=3.221$), MACS J0416.1$-$2403 ID16 ($z=2.095$), and MACS J1206.2$-$0847 ID14 ($z=3.753$). Eight multiple images were observed for the first SL system, and six for the latter two. We focused on the main deflector of each galaxy-scale SL system (identified as members 8971, 8785, and 3910, respectively), and modelled its total mass distribution with a truncated isothermal sphere. To account for the lensing effects of the remaining components of the cluster, we took the most accurate SL model of its mass distribution available. To include the uncertainty and the systematics affecting the cluster-scale mass models, we explored the posterior probability distribution of its parameters and extracted 100 cluster mass distributions. For each of them, we optimised the mass parameters of the galaxy-scale lens: the bootstrapping procedure allowed us to obtain a realistic estimate of the uncertainty on their values.}
   {We measured a truncation radius value of $6.1^{+2.3}_{-1.1} \, \mathrm{kpc}$, $4.0^{+0.6}_{-0.4} \, \mathrm{kpc}$, and $5.2^{+1.3}_{-1.1} \, \mathrm{kpc}$ for members 8971, 8785, and 3910, corresponding to total mass values of $M=1.2^{+0.3}_{-0.1}\times10^{11} \, M_\odot$, $M=1.0^{+0.2}_{-0.1} \times 10^{10} \, M_\odot$, and $M=6.3^{+1.0}_{-1.1} \times 10^{10} \, M_\odot$, respectively. Alternative non-truncated models with a higher number of free parameters do not lead to an improved description of the SL system and show some parametric degeneracies. We measured the stellar-to-total mass fraction within the effective radius for the three cluster members, finding $0.51\pm0.21$, $1.0\pm0.4$, and $0.39\pm0.16$, respectively.}
   {We find that a parameterisation of the physical properties of cluster galaxies in SL models based on power-law scaling relations with respect to the observed total luminosity cannot accurately describe the compactness of the members over their full total mass range. Our results, instead, agree with recent modelling of the cluster members based on the Fundamental Plane relation. Finally, we report good agreement between our predicted values of the stellar-to-total mass fraction within the effective radius and those of early-type galaxies from the Sloan Lens ACS Survey. Our work significantly extends the regimes of the current samples of lens galaxies, towards the mass range that will be probed by the \textit{Euclid}, \textit{Rubin}, and \textit{James Webb} Telescopes.}

   \keywords{gravitational lensing: strong --
                galaxies: clusters: general --
                galaxies: clusters: individual (MACS J0416.1$-$2403) --
                galaxies: clusters: individual (MACS J1206.2$-$0847) -- dark matter -- cosmology: observations
               }

   \maketitle
%
\section{Introduction}\label{s1}

Gravitational lensing has recently become an extremely effective technique to study the dark-matter (DM) distribution in galaxies and clusters of galaxies \citep[e.g.][]{natarajan97,treu10}. The observed light deflection effects only depend on the total gravitational potential of the lens, without any discrimination between the baryonic and DM components. Due to their high mass and deep gravitational potential well, clusters often determine strong lensing (SL) of several tens of background sources. Strong lensing studies have led to measurements of the cluster total mass profile with an uncertainty of a few percent near the core, that is within a few hundreds of kiloparsecs from the cluster centre \citep[e.g.][]{grillo15,jauzac15,caminha17a,caminha17b,sharon20,acebron22}. The mass budget of galaxy clusters is made up by DM for up to more than $85\%$ of the total mass, while the remaining, baryonic, mass component is dominated by a hot plasma, the intra-cluster medium (ICM), whose mass distribution can be estimated from X-ray data. Observations of SL can thus be combined with baryonic mass diagnostics to disentangle the number and mass distributions of the DM haloes of galaxy clusters from the total mass profile of the lens \citep[e.g.][]{annunziatella17,sartoris20}.
   
Several \textit{Hubble} Space Telescope (HST) photometric campaigns have supported the effort to identify multiply imaged sources in the cores of different samples of galaxy clusters, in order to build detailed SL models. The most notable examples are the Cluster Lensing And Supernova survey with \textit{Hubble} \citep[CLASH,][]{postman12}, the \textit{Hubble} Frontier Fields \citep[HFF,][]{lotz17} programme, the Reionization Lensing Cluster Survey \citep[RELICS,][]{coe19}, and the Beyond Ultra-deep Frontier Fields And Legacy Observations \citep[BUFFALO,][]{steinhardt20}. Spectroscopic follow-up campaigns on the Very Large Telescope (VLT), carried out with multi-object spectrographs, such as CLASH-VLT \citep[dark matter mass distributions of \textit{Hubble} treasury clusters and the foundations of LCDM structure formation models,][]{rosati14}, complemented with data from the integral-field Multi Unit Spectroscopic Explorer \citep[MUSE,][]{bacon10}, have identified and confirmed up to more than 1000 cluster members \citep[e.g.][]{mercurio21} and more than 200 multiple images per cluster \citep[e.g.][hereafter B23]{bergamini22}. Thanks to these data, lensing models have reached extremely high resolution in mapping the mass distribution of all the cluster components, down to the scale of the single member galaxies.
   
In the framework of the currently adopted \textrm{$\Lambda$}CDM cosmological model, dominated by CDM and with a cosmological constant $\Lambda$, cosmological simulations allow us to describe the formation and evolution of DM haloes at different scales. They show that haloes form hierarchically from subsequent mergers of smaller structures \citep[e.g.][]{tormen97,moore99,borgani11}: numerous less massive haloes, or sub-haloes, are therefore found in the proximity of the most massive ones. The accuracy of simulations has been significantly increased in the last few years by the inclusion of baryons and of the physical effects of their interplay with DM. Simulations can therefore be used to obtain quantitative predictions on the expected number and mass distributions of haloes and sub-haloes in the Universe. Any significant discrepancy between these results and what is inferred from observations may imply that the formation of structures does not proceed as forecast by the underlying cosmological hypotheses and therefore put into question the \textrm{$\Lambda$}CDM paradigm and/or our current understanding of the effects of baryons in shaping the mass distribution of the DM haloes.
   
Comparing 25 clusters extracted from a suite of \textrm{$\Lambda$}CDM hydrodynamical simulations \citep[][]{planelles14,rasia15} with 11 state-of-the-art SL models of massive clusters observed in CLASH, \citet{meneghetti20} reported a discrepancy of around an order of magnitude between the simulated and observed probability for the clusters to produce galaxy-galaxy strong lensing (GGSL) events (i.e. SL phenomena in which several multiple images are observed around one or a few cluster members). This discrepancy can be interpreted as a result of simulated galaxy cluster members being less compact than their observed counterparts. To investigate the possible impact of the numerical setup chosen for the simulations, \citet{meneghetti22}, \citet{ragagnin22}, and \citet{meneghetti23} repeated the test by \citet{meneghetti20} with different simulation resolutions and baryonic feedback schemes, confirming the previously reported tension with observations. Focusing, instead, on the possible systematics affecting SL models, \citet{granata22}, hereafter G22, tested the impact on the discrepancy of the power-law scaling relations used in lensing models to link the mass of the cluster members to their luminosity, replacing them with the Fundamental Plane \citep[FP,][]{dressler87,djorgovski87,bender92} relation, defined in the three-dimensional ($\log \sigma_0, \, \log R_e, \, \mathrm{SB}_e$) space, where $\sigma_0$ is the central velocity dispersion, $R_e$ is the effective radius, and $\mathrm{SB}_e$ is the average surface brightness within $R_e$ in $\mathrm{mag \, arcsec^{-2}}$ units. The procedure allows for a more complex description of the physical properties of the members, but does not significantly reduce the observed discrepancy.
   
In this work, we considered the reference sample of lens clusters included in \citet{meneghetti20}: Abell S1063 (AS1063), at $z=0.348$, MACS J0416.1$-$2403 (MACS J0416), at $z=0.396$, and MACS J1206.2$-$0847 (MACS J1206), at $z=0.439$. The first two clusters were part of the HFF sample, while all three were CLASH and CLASH-VLT targets. We examined all main GGSL events in the three clusters, looking for systems with a clear morphology, in which several multiple images are observed very close to a cluster member, providing us with stringent constraint on its total mass profile, and therefore on its compactness. We selected two GGSL systems in MACS J0416, identified as ID14 and ID16 in \citetalias{bergamini22}, and one in MACS J1206, identified as ID14 in \citetalias{bergamini19}. The  first analysis of this system was performed by \citet{grillo14}.
   
Gravitational lensing, in combination with stellar dynamics, has allowed for a considerable improvement of our understanding of the internal structure of galaxies, such as their stellar-to-total mass fraction and the mass density distribution of their DM haloes \citep[see ][]{shajib22}. In particular, observational campaigns such as the Lenses Structure and Dynamics \citep[LSD;][]{treu04}, the Sloan Lens ACS Survey \citep[SLACS;][]{bolton06, treu06, auger10}, the Strong Lensing Legacy Survey \citep[SL2S;][]{gavazzi12}, and the Dark Energy Survey \citep[DES;][]{abbott19} have mostly focused on isolated, massive early-type galaxies, due to observational limitations. Their total mass distribution is very well fit by an isothermal profile out to a large distance from their centre \citep{treu06}. In the next few years, new, wide, and deep surveys such as \textit{Euclid} and the \textit{Rubin} Observatory Legacy Survey of Space and Time (LSST) will boost the number of known galaxy-scale lenses by a few orders of magnitude, and significantly extend the lower mass threshold of observed lenses \citep{collett15}.

The intense gravitational field of the cluster leads to secondary critical lines forming around several faint galaxies. In this work, we build SL models for galaxies whose mass is too low for current surveys of lens galaxies to detect them in the field, probing a total lens mass range which will only be fully explored by the upcoming surveys. Furthermore, SL models of clusters favour truncated mass profiles for the cluster members, with a half-mass radius a few times larger than the half-light radius (\citetalias{granata22}). In this work, we evaluate the robustness of the assumption of truncated mass profiles for the cluster members and test possible alternative parametrisations.
   
The paper is organised as follows. In Sect. \ref{s2}, we give details on the reference SL models for MACS J0416 and MACS J1206, published in \citetalias{bergamini22} and \citetalias{bergamini19}, respectively. In Sect. \ref{s3}, we present our observations and the layout of the three GGSL systems. In Sect. \ref{s4}, we build a SL model for the lens galaxies, and in the following Sect. \ref{s5}, we discuss their inferred properties, with a special focus on their compactness. Finally, in Sect. \ref{s6}, we summarise our results. In this work, we use a flat $\mathrm{\Lambda}$CDM cosmology with $\Omega_\mathrm{m}=0.3$ and $H_0 = 70$ $\mathrm{km \, s^{-1} \, Mpc^{-1}}$, in which $1''$ corresponds to a scale of 5.34 kpc at $z=0.396$, the redshift of MACS J0416, and of 5.68 kpc at $z=0.439$, the redshift of MACS J1206. All magnitudes are expressed in the AB system.
   
\section{Reference strong lensing models}\label{s2}

As anticipated, we considered the most recent versions of the SL models of MACS J0416 and MACS J1206 adopted in \citet{meneghetti20}. They were presented in \citetalias{bergamini22} and \citetalias{bergamini19}, respectively. Both models were built using the publicly available code \texttt{LensTool} \citep[][]{kneib96,jullo07,jullo09}. All mass components were described with a parametric dual pseudo-isothermal elliptical (dPIE) mass density profile \citep[][]{limousin05,eliasdottir07}, which is the ellipsoidal generalisation of a truncated isothermal sphere with a central core. The three-dimensional mass density profile of a spherical dPIE is \citep{eliasdottir07}
\begin{equation}\label{dpieeq}
    \rho(r)=\frac{\rho_0}{(1+r^2/r_c^2)(1+r^2/r_t^2)},
\end{equation}
where $r$ is the distance from the halo centre, while $r_c$ and $r_t$ are the core and the truncation radii, respectively. Equation (\ref{dpieeq}) implies a smooth transition for the mass density between a central flat core for $r<r_c$, an isothermal behaviour ($\rho(r)\propto r^{-2}$) for $r_c<r<r_t$, and a steeper profile ($\rho(r)\propto r^{-4}$) for $r>r_t$. The truncation radius can also be interpreted as the three-dimensional half-mass radius \citepalias[][]{bergamini19}. The central density scale, $\rho_0$, is related to the isothermal velocity dispersion $\sigma$ by \citep{limousin05}
\begin{equation}\label{sigmalensing}
   \rho_0 = \frac{\sigma^2}{2 \pi G} \frac{r_c+r_t}{r_c^2 r_t}. 
\end{equation}
The velocity dispersion parameter $\sigma$ is connected to the observed aperture-averaged stellar velocity dispersion by a projection coefficient (see Appendix C of \citetalias{bergamini19}). The ellipticity is introduced substituting the projected distance from the centre $R$ with $\hat{R}$ such that \citep{eliasdottir07}
\begin{equation} \label{ellvalue}
    \hat{R}^2=\frac{x^2}{\left(\frac{2a}{a+b}\right)^2}+\frac{y^2}{\left(\frac{2b}{a+b}\right)^2},
\end{equation}
where $a$ and $b$ are the major and minor projected semi-axes of the ellipsoid, and $x$ and $y$ are the coordinates along them\footnote{Equation (\ref{ellvalue}) implies that, compared to the spherical case, the area enclosed by a given iso-surface-density contour changes by a factor $(4ab)(a+b)^{-2}$.}.

The diffuse DM haloes and the ICM were modelled with cluster-scale dPIEs. The total mass distributions of each cluster member was described with a spherical coreless dPIE. An external shear term was taken into account while modelling MACS J1206. Following \citet{bonamigo18}, the ICM mass distribution was fixed from X-ray observations, while all remaining components were optimised.
The optimisation of the free parameters in \texttt{LensTool} is driven by a $\chi^2$-based likelihood which quantifies how well a given set of parameters reproduces the observed positions of the multiple images. The function can be written as
\begin{equation} \label{chi2eq}
\chi^2 (\boldsymbol{\theta}) = \sum^{N_{\mathrm{fam}}}_{j=1}\sum^{N^j_{\mathrm{img}}}_{i=1} \left( \frac{\lVert \mathbf{x}_{\mathrm{obs}\,i,j} - \mathbf{x}_{\mathrm{pred}\,i,j}(\boldsymbol{\theta})\lVert}{\sigma_{x \, i,j}} \right)^2,
\end{equation}
where $\mathbf{x}_{\mathrm{obs} \, i,j}$ and $\sigma_{x \, i,j}$ are the observed position of the $i$-th image of the $j$-th family and its uncertainty, respectively, and $\mathbf{x}_{\mathrm{pred}\,i,j}(\boldsymbol{\theta})$ is the position of the same image as predicted by the model defined by the set of parameter values $\boldsymbol{\theta}$. The accuracy of a given model at reproducing a given set of multiple images is usually measured with the root mean square difference between their model-predicted and observed positions (indicated as $\Delta_\mathrm{rms}$).

A large sample of multiple images is therefore crucial to increase the number of observational constraints for the determination of the best-fit set of parameters. The SL model of MACS J0416 presented in \citetalias{bergamini22} is based on the largest sample of spectroscopically confirmed multiple images ever built for such scope: 237 multiple images from 88 background sources in the redshift range $z = 0.94 - 6.63$. The best-fit model has $\Delta_\mathrm{rms}=0.43''$ and includes 213 cluster members.
The SL model of MACS J1206 from \citetalias{bergamini19}, on the other hand, includes 82 multiple images from 27 background sources in the redshift range $z = 1.01 - 6.01$. The best-fit model has $\Delta_\mathrm{rms}=0.46''$ and includes 258 cluster members.
The values of $\Delta_\mathrm{rms}$ found by \citetalias{bergamini22} and \citetalias{bergamini19}, summarised in Table \ref{refmodels}, are significantly higher than the astrometric uncertainty on the positions of multiple images (typically smaller than $0.01''$). They are however in line with the state of the art of parametric SL models of massive clusters: in spite of the availability of exquisite samples of spectroscopic multiple images, models are still affected by systematics that have an impact on their predicted image positions, such as the choice of the parametrisation of the cluster total mass distribution and the lensing effects of the cluster environment \citep{meneghetti17,acebron17,grillo16,johnson16}. For instance, as found by \citet{chirivi18}, the line-of-sight mass distribution of MACS J0416, which is not included in \citetalias{bergamini22}, has a significant impact on the reconstruction of the multiple images, although its inclusion is not sufficient to reconcile observations and model predictions. Galaxy-scale SL models are not as affected by intrinsic systematics: we need to fully propagate the uncertainty impacting the cluster-scale mass distribution on the determination of the galaxy-scale parameters.

\subsection{Modelling the mass distribution of the cluster members}

As described in the previous sub-sections, the cluster members of MACS J0416 and MACS J1206 were modelled with spherical dPIEs with a vanishing core radius. Their total mass only depends on two parameters: their velocity dispersion parameter $\sigma$, and their truncation radius $r_t$. In the case of a vanishing core radius, the velocity dispersion parameter $\sigma$ of a dPIE profile, defined in Eq. (\ref{sigmalensing}), is well approximated by the measured stellar central velocity dispersion $\sigma_0$ \citepalias[][]{bergamini19}. In order to reduce the number of free parameters during the model optimisation, power-law scaling relations were calibrated between the two free parameters and the total luminosity of the cluster members. These relations are usually expressed as
\begin{equation}\label{scaling1}
       \sigma_{i} = \sigma^\mathrm{ref} \left(\frac{L_i}{L_0}\right)^{\alpha}, \\
\end{equation}
\begin{equation}\label{scaling2}
       r_{t,i}= r^\mathrm{ref}_t \left(\frac{L_i}{L_0}\right)^{\beta}, \\
\end{equation}
where $L_0$ is the luminosity of the BCG. Cluster-scale SL models are mostly sensitive to the total mass of the cluster members, rather than to the way it is distributed on the scale of the single galaxy, determining a clear degeneracy between $\sigma$ and $r_t$. 

An important step forward in the reduction of the impact of this degeneracy has been driven by the MUSE integral-field spectroscopy, which allows for the measurement of the velocity dispersion of several cluster members, thus obtaining an independent (i.e. not related to the SL constraints) observational prior on the values of $\alpha$ and $\sigma^\mathrm{ref}$, the slope and the normalisation of the first scaling law presented in Eq. (\ref{scaling1}) \citep[also known as the Faber-Jackson law,][]{faber76}. Respectively 64 and 58 measured velocity dispersions have been included in the SL models of MACS J0416 and MACS J1206. This procedure partly breaks the degeneracy between the two parameters describing the cluster members. Using the SL forward modelling code \texttt{GravityFM} (Bergamini et al. in preparation), \citetalias{bergamini22} reconstructed the original surface brightness profile of some sources lensed by MACS J0416 and then traced back their images to the lens plane. They showed that, despite the parametric degeneracies involved, SL models can faithfully reproduce the observed morphology of the GGSL images, and that the accuracy of the reconstruction is enhanced by the inclusion of a larger sample of observed multiple images as a constraint for the optimisation of the model. The degeneracy between the values of the velocity dispersion and of the truncation radius of the cluster members has also been addressed by \citetalias{granata22} by replacing the Faber-Jackson law from Eq. (\ref{scaling1}) with a newly-calibrated FP, a more complex scaling law of which the Faber-Jackson is a projection. This allowed to estimate the central velocity dispersion for all the cluster members included in the model from their observed magnitude and half-light radius.

\begin{table}
\caption{Reference lensing models of MACS J0416 (\citetalias{bergamini22}) and MACS J1206 (\citetalias{bergamini19}): relevant parameters. $N_\mathrm{m}$ and $N_\mathrm{m}^\mathrm{s}$ indicate the number of cluster members included in the model and the number of those with measured velocity dispersion, respectively. $N_\mathrm{i}$ and $N_\mathrm{s}$ indicate the number of multiple images included in the model and the number of sources, respectively.}             
\label{refmodels}      
\centering                          
\begin{tabular}{c c c c}        
\hline \noalign{\smallskip}                
Cluster &  $N_\mathrm{m}\,(N_\mathrm{m}^\mathrm{s})$ & $N_\mathrm{i}\,(N_\mathrm{s})$ & $\Delta_\mathrm{rms}$ \\    
\noalign{\smallskip} \hline  \noalign{\smallskip}                      
            MACS J0416 & $213\,(64)$ & $237\,(88)$ & $0.43''$   \\
            MACS J1206 & $258\,(58)$ & $82\,(27)$ & $0.46''$ \\
 \noalign{\smallskip} \hline                                   
\end{tabular}
\end{table}

The $\sigma-r_t$ degeneracy has a very significant impact on the estimate of the compactness of the cluster members. The same total mass value for the cluster members can be obtained with different combinations of the two parameters resulting in a more (or less) compact mass distribution. On the other hand, in a GGSL system, the positions of the multiple images are stringent constraints on the critical lines around the lens galaxies, which strongly depend on its compactness, thus breaking the degeneracy.

\section{The galaxy-galaxy strong lensing systems}\label{s3}

In this section, we lay out the structure of the three GGSL systems on which we focus our work, the data available for the cluster members, and the procedure followed to obtain the multiple image identification which we later adopt in our lensing analyses. 

\subsection{MACS J0416.1$-$2403 ID14}\label{ss31}

\begin{figure}
   \centering
   \includegraphics[width=9cm]{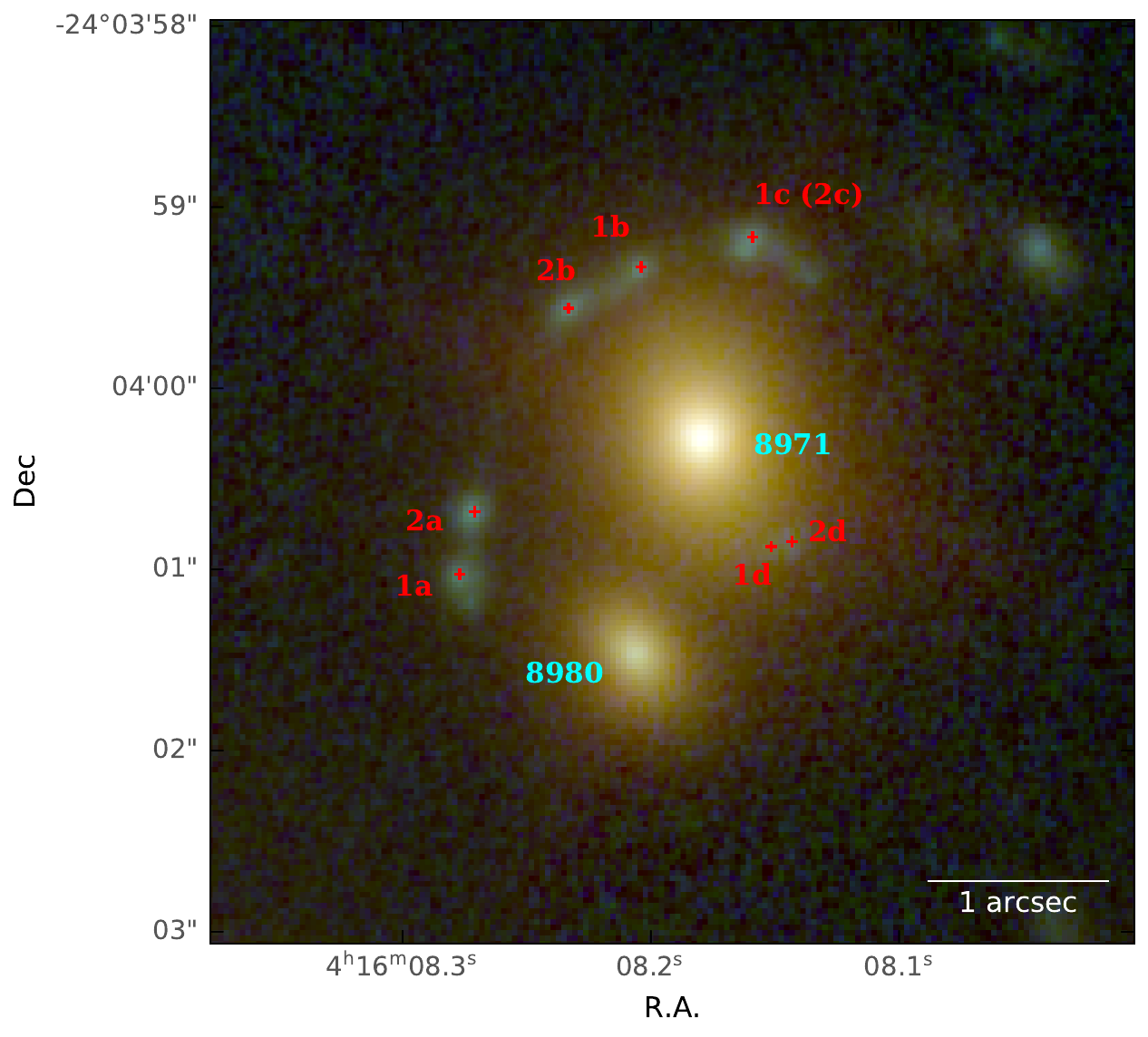}
      \caption{\textit{Hubble} Frontier Fields RGB image of MACS J0416.1$-$2403 ID14. The two cluster members are identified as in \citetalias{bergamini22}. The two components of the source are identified as 1 and 2. The four multiple images are indicated with the letters a, b, c, and d.}
         \label{id14}
   \end{figure}

The source ID14, studied in detail by \citet{vanzella17}, is composed of a pair of faint, young, and compact stellar systems. It has been included in \citetalias{bergamini22} with a spectroscopic redshift of $z=3.221$ \citep{balestra16}. It is split by the gravitational potential of MACS J0416 into three images separated by up to $50''$. One of the three images, shown in Fig. \ref{id14}, falls close to a pair of elliptical cluster galaxies (members 8971 and 8980 in the reference SL model) and is further split into four images. Following \citet{vanzella17}, we refer to the two sources as 1 and 2, and to the four images as a, b, c, and d. Knots 1 and 2 have a very similar magnitude in images a and b. Image 1c is the brightest observed, while 2c is much fainter and its position is hard to measure precisely. A detailed photometric study of the lensing system has been performed to build the most recent lensing models: we therefore adopted the positions thereby reported for the images a, b, and c. As such, we also chose to consider 2c in the same position as 1c for the purpose of SL modelling, with a larger uncertainty on its value. In the case of image d, unlike \citetalias{bergamini22}, where only image 1d was included in the model, we identified the two components of the source and measured their positions separately. The adopted multiple-image positions are reported in Table \ref{id14img}.

\begin{table}
\caption{Multiple image positions for the GGSL system MACS J0416 ID14. We report the relative positions of the multiple images with respect to the centre of the cluster member 8971, for which we provide the values of R.A. and Dec. The images are identified as in Fig. \ref{id14}. We also report the position of the centre of the cluster member 8980.}             
\label{id14img}      
\centering                          
\begin{tabular}{c c c}        
\hline \noalign{\smallskip}                
Reference & $64.034084$ & $-24.066743$ \\
\noalign{\smallskip} \hline  \noalign{\smallskip}               
Position &  $x \, ('')$ & $y \, ('')$ \\    
\noalign{\smallskip} \hline  \noalign{\smallskip}                      
            1a & $-1.34$ & $-0.77$    \\
            2a & $-1.26$ & $-0.42$ \\
            1b & $-0.34$ & $0.93$ \\
            2b & $-0.74$ & $0.70$ \\
            1c (2c) & $0.27$ & $1.09$ \\
            1d & $0.49$ & $-0.59$ \\
            2d & $0.38$ & $-0.57$ \\
\noalign{\smallskip} \hline  \noalign{\smallskip} 
            8980 & $-0.36$ & $-1.20$ \\
 \noalign{\smallskip} \hline                                   
\end{tabular}
\end{table}   

\subsection{MACS J0416.1$-$2403 ID16} \label{ss32}

\begin{figure}
   \centering
   \includegraphics[width=9cm]{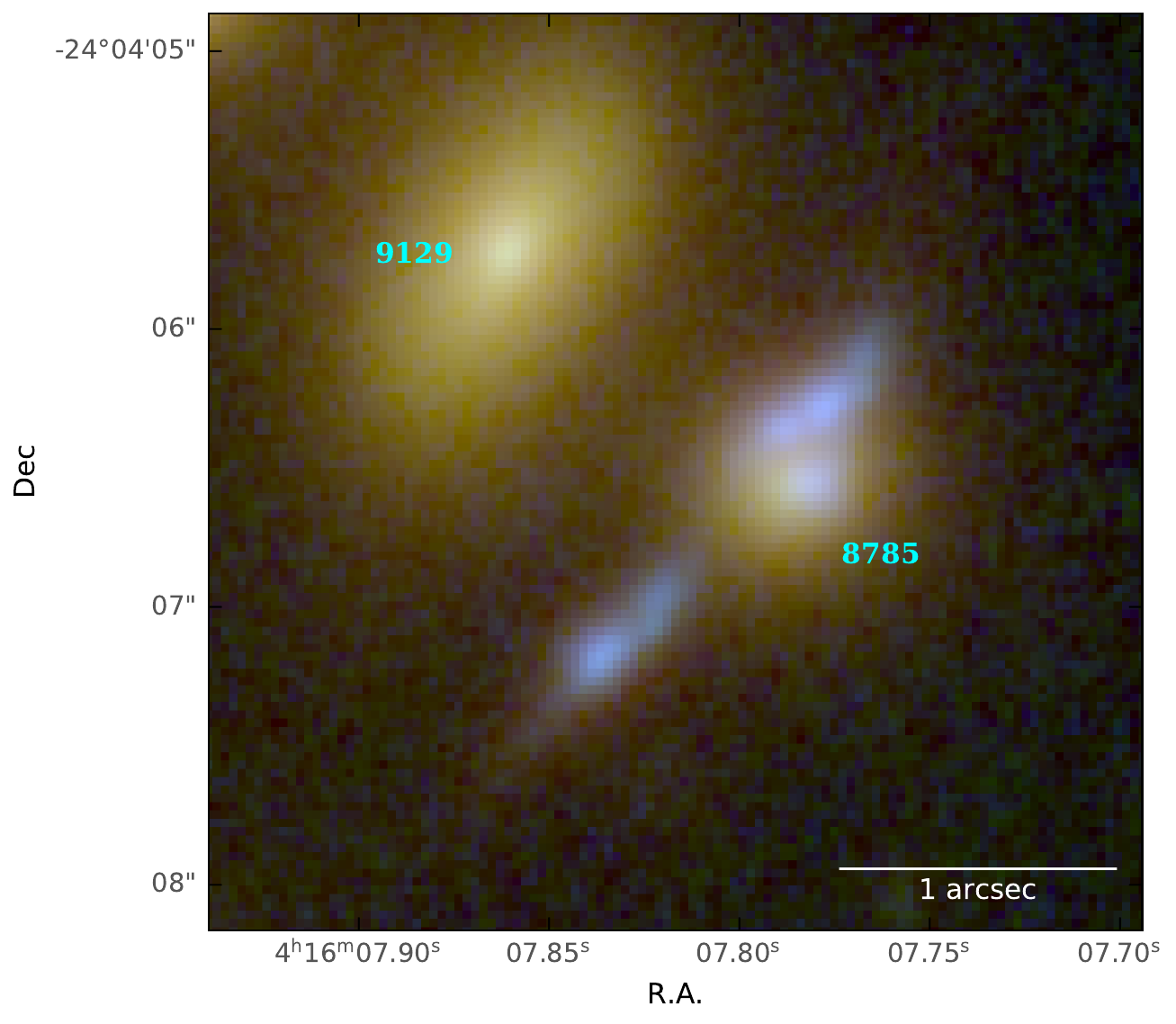}
      \caption{\textit{Hubble} Frontier Fields RGB image of MACS J0416.1$-$2403 ID16. The two cluster members are identified as in \citetalias{bergamini22}.}
         \label{id16}
\end{figure}

\begin{figure*}
  \centering
\includegraphics[scale=0.39] {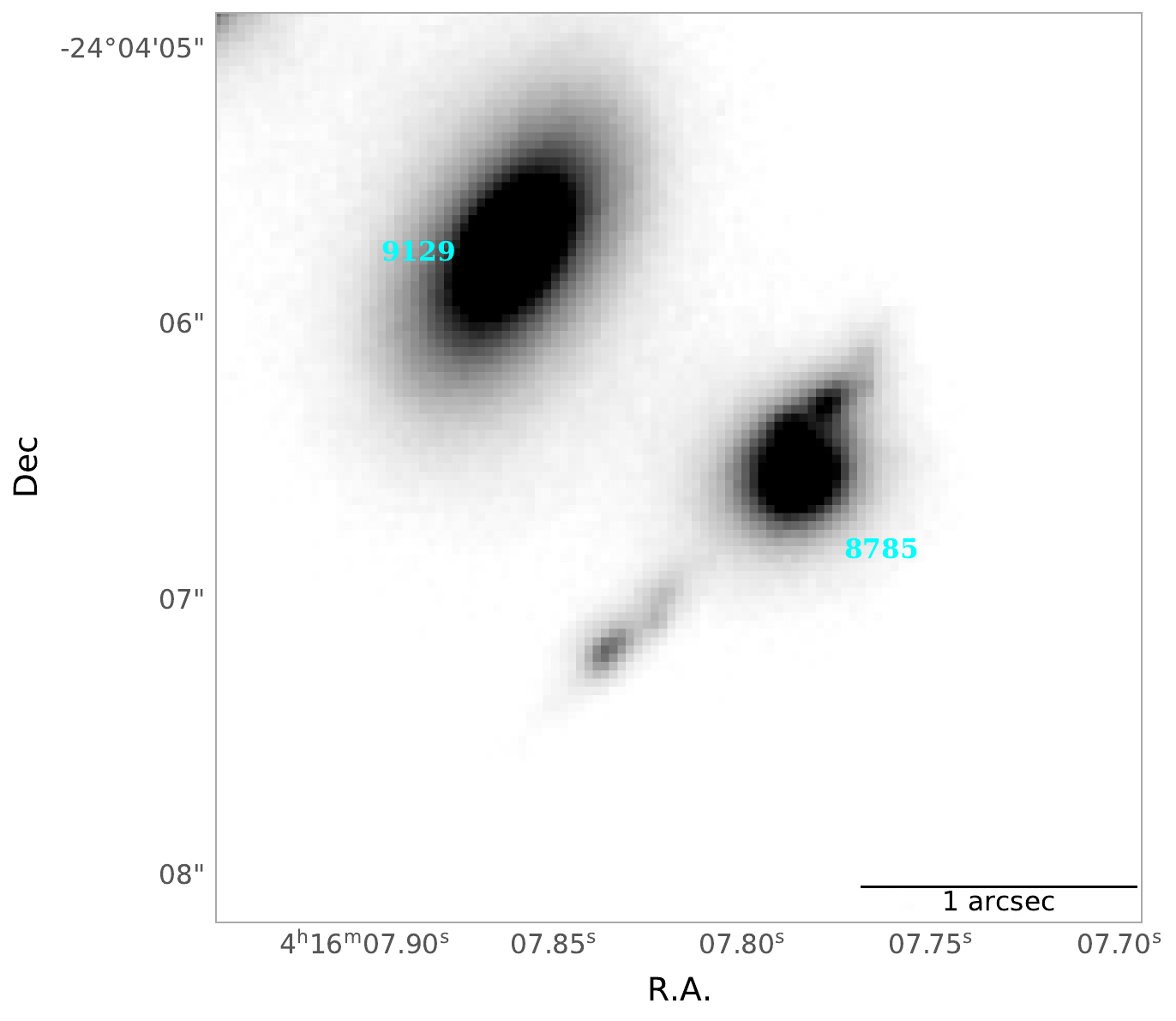} 
\includegraphics[scale=0.39] {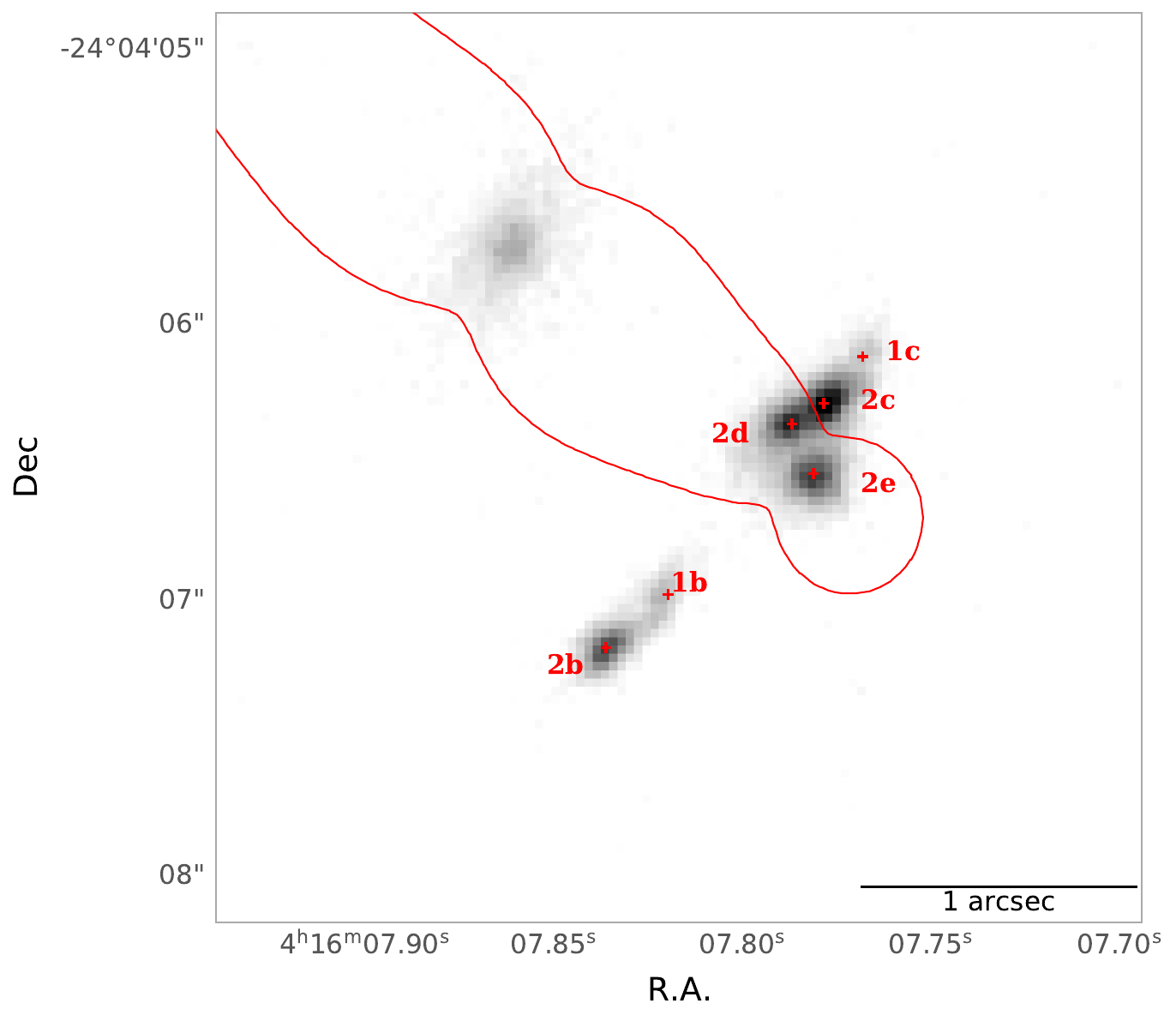}

  \caption{\textit{Hubble} Frontier Fields imaging of MACS J0416.1$-$2403 ID16. Left panel: the two cluster members in the ACS F814W band. Right panel: the multiple images in the ACS F435W band. The two components of the source are identified as 1 and 2. The four multiple images are indicated with the letters b, c, d, and e. The critical lines predicted by the model by \citetalias{bergamini22} for a source at redshift $z=2.095$ (the same as ID16) are marked in red.}
  \label{crit}
  \end{figure*}

The double source identified as ID16 in \citet{bergamini21}, at a spectroscopic redshift of $z=2.095$ \citep{balestra16}, is split into three multiple images by the gravitational potential of the cluster main halo. One of the three images falls very close to a pair of elliptical cluster members (8785 and 9129 in the reference SL model, see Fig. \ref{id16}). The light distribution in the bluest of the HFF bands (ACS F435W) suggests that two further multiple images might fall very close to the centre of cluster member 8785, one of them being almost completely superimposed to it. In order to correctly measure the position of this image and include it in our lensing model, we therefore separated its light from that of the cluster galaxy. We did this in the HFF ACS F606W band, in which both the cluster member and the multiple image are visible. To perform this task, first, we masked the light from the lensed image. The masked pixels are determined in the HST F435W band (where ID16 is less affected by the light of the cluster member). This way, we could run \texttt{GALFIT} \citep{peng10} on the masked F606W image to model the light of the cluster member and measure its structural parameters. In the \texttt{GALFIT} run, the position of the cluster member and its S\'ersic index $n=0.5$ (Gaussian light distribution) were kept fixed. Parameters such as the magnitude and the half-light radius of the cluster member were manually tuned and fixed to minimise the residual image obtained after its subtraction. No significant over- or under-subtraction regions were observed in the residual image and the residual values are less than 20\% of the original image value in every pixel. We therefore used the residual F606W image to measure the positions of all the multiple images, determining their $x$ and $y$ coordinates with Gaussian fits of the light profile. We refer to the two sources as 1 and 2 and to the three images produced by the cluster member 8785 as c, d, and e. We also included image b in the model, which is only at a distance of around $1''$ from the cluster galaxy, and for which we have also measured the positions of the two components. The two sources are not clearly resolved in images d and e: we only included the brightest source, identified as 2, in our model. In Fig. \ref{crit}, we confirm the new identification of two additional multiple images using the reference cluster-scale SL model. Even when only images a, b, and c are included in the optimisation process, the critical lines close to cluster member 8785 create images d and e. As shown in \citetalias{bergamini22}, SL forward modelling of the source ID16 performed with \texttt{GravityFM} allows us to very accurately reproduce the observed multiple image configuration and surface brightness distribution. As such, our new multiple image catalogue, presented in Table \ref{id16img}, has been adopted by \citetalias{bergamini22}.

\begin{table}
\caption{Multiple image positions determined from HFF photometry for the GGSL system MACS J0416 ID16. We report the relative positions of the multiple images with respect to the centre of the cluster member 8785, for which we provide the values of R.A. and Dec. The images are identified as in Fig. \ref{id16}. We also report the position of the centre of the cluster member 9129.}             
\label{id16img}      
\centering                          
\begin{tabular}{c c c}        
\hline \noalign{\smallskip}                
Reference & $64.032442$ & $-24.068485$ \\
\noalign{\smallskip} \hline  \noalign{\smallskip}
Position &  $x \, ('')$ & $y \, ('')$ \\    
\noalign{\smallskip} \hline  \noalign{\smallskip}                      
            1b & $-0.51$ & $-0.44$    \\
            2b & $-0.70$ & $-0.64$ \\
            1c & $0.23$ & $0.42$ \\
            2c & $0.09$ & $0.25$ \\
            2d & $-0.03$ & $0.18$ \\
            2e & $0.05$ & $-0.004$ \\
\noalign{\smallskip} \hline  \noalign{\smallskip} 
            9129 & $-1.02$ & $0.82$ \\
 \noalign{\smallskip} \hline                                   
\end{tabular}
\end{table}

\subsection{MACS J1206.2$-$0847 ID14}\label{ss33}

\begin{figure}
   \centering
   \includegraphics[width=9cm]{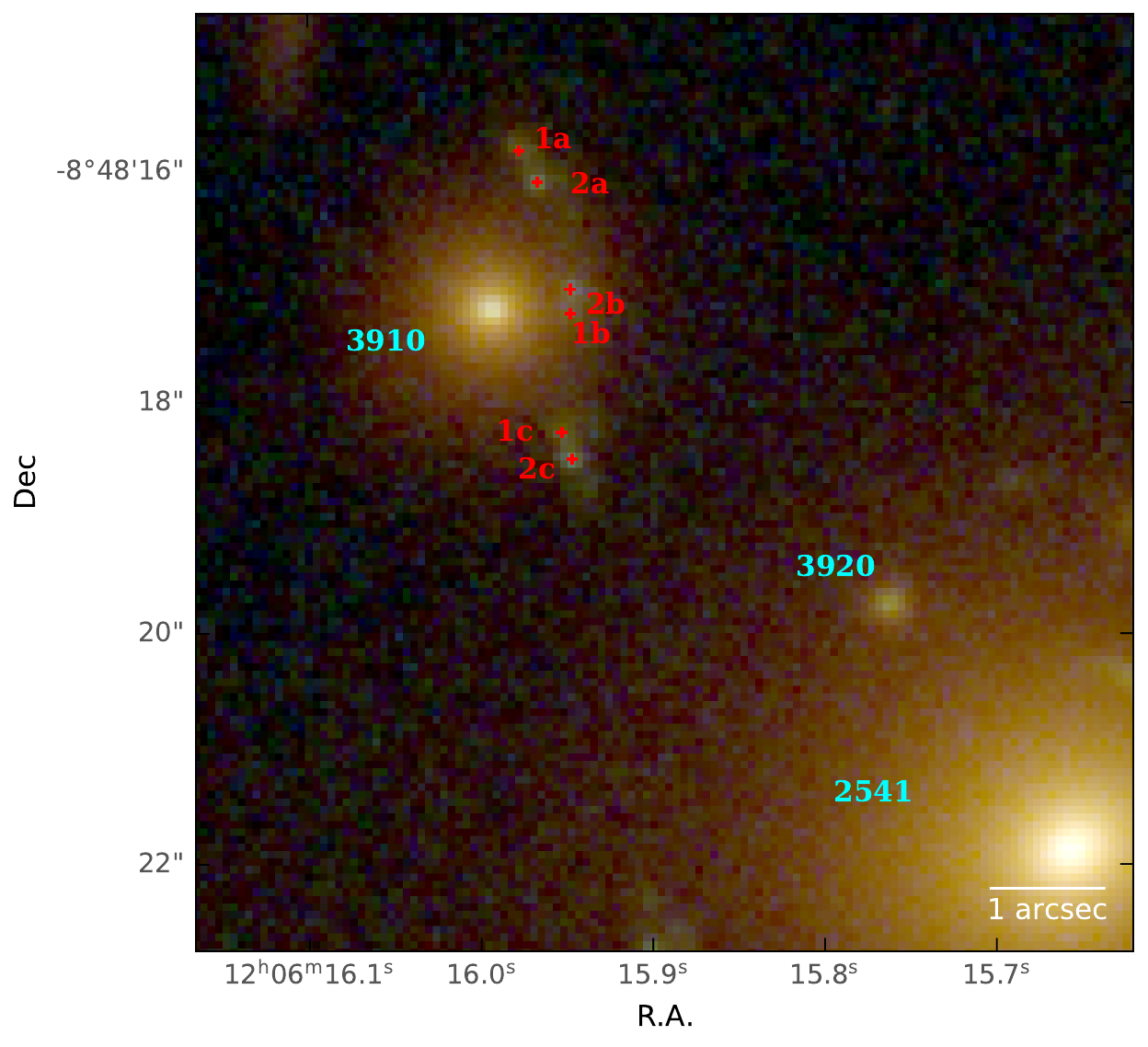}
      \caption{CLASH RGB image of MACS J1206.2$-$0847 ID14. The two cluster members are identified as in \citetalias{bergamini19}. The two components of the source are identified as 1 and 2. The three multiple images are indicated with the letters a, b, and c.}
         \label{id141206}
\end{figure}

The double source ID14 was first included by \citet{grillo14} in a similar galaxy-scale lensing study, before a cluster-scale lensing model for MACS J1206 was available. It is treated by \citetalias{bergamini19} as a single source, with a spectroscopic redshift of $z=3.753$ \citep{biviano13,caminha17b}. Five multiple images are observed, three of them very close to the centre of a cluster member (ID 3910), on which we focus our attention. Their positions are also influenced by the deflection caused by the second brightest cluster member (ID 2541). We refer to the two sources as 1 and 2 and to the three images as a, b, and c, as shown in Fig. \ref{id141206}. We measured their $x$ and $y$ positions, reported in Table \ref{id141206img}, using CLASH photometry in the ACS F435W band.

\begin{table}
\caption{Multiple image positions determined from CLASH photometry for the GGSL system MACS J1206 ID14. We report the relative positions of the multiple images with respect to the centre of the cluster member 3910, for which we provide the values of R.A. and Dec. The images are identified as in Fig. \ref{id141206}. We also report the position of the centre of the cluster member 2541.}             
\label{id141206img}      
\centering                          
\begin{tabular}{c c c}        
\hline \noalign{\smallskip}                
Reference & $181.566661$ & $-8.804784$ \\
\noalign{\smallskip} \hline  \noalign{\smallskip}
Position &  $x \, ('')$ & $y \, ('')$ \\    
\noalign{\smallskip} \hline  \noalign{\smallskip}                      
            1a & $0.26$ & $1.36$    \\
            2a & $0.40$ & $1.10$ \\
            1b & $0.68$ & $-0.04$ \\
            2b & $0.68$ & $0.18$ \\
            1c & $0.61$ & $-1.06$ \\
            2c & $0.70$ & $-1.29$ \\
\noalign{\smallskip} \hline  \noalign{\smallskip} 
            2541 & $5.02$ & $-4.67$ \\
 \noalign{\smallskip} \hline                                   
\end{tabular}
\end{table}

\section{Strong lensing modelling}\label{s4}

Our main aim is to use galaxy-scale SL modelling to directly constrain the compactness of the cluster members, and to compare our results with those obtained with cluster-scale modelling, to study their effectiveness at recovering the mass distribution of the cluster sub-structures. In previous galaxy-scale SL studies in clusters \citep[e.g.][]{grillo14,parry16}, a simplified description of the cluster-scale mass distribution was included in the SL models to account for its effects on the image deflection. An incorrect determination of the cluster-scale mass distribution can significantly hinder the description of galaxy-scale lenses, especially as far as their azimuthal structure is concerned. In our case, instead, the cluster-scale mass distribution is constrained with accuracy, owing to SL models based on up to more than 200 multiple images of background sources.

In SL models of massive galaxy clusters, the diffuse and galaxy-scale mass components are jointly optimised with the same set of constraints, as described in Sect. \ref{s2}. The cluster-scale DM haloes dominate the total mass budget and determine the position of its primary critical lines, therefore the parameters that describe their mass distribution have the largest impact on the positions of the multiple images \citep[][]{meneghetti17,limousin22}. In addition, the scaling laws connecting the parameters adopted to model the cluster galaxies are mostly influenced by the properties of the most massive members, and power-laws can be too simple to describe the wide mass and morphology range of the cluster members \citep[see \citetalias{granata22},][]{beauchesne23}. The total mass distribution of the low- and intermediate-mass cluster galaxies is therefore not entirely constrained by cluster-scale SL models, as shown by the difference between the results obtained by describing them with different scaling laws \citepalias[][]{granata22}. When modelling a galaxy-scale lens, we thus only optimised the parameters defining the cluster galaxies whose mass distribution significantly impacts the positions of the multiple images, keeping fixed the mass distribution of the rest of the galaxy cluster. To build the $\chi^2$ function driving the parametric optimisation (see Sect. \ref{s2}), we only included the multiple images close to the main galaxy-scale lens (within $2''$ from its centre). More distant multiple images could provide us with additional constraints the position of the source, but their position is primarily determined by the mass distribution of other cluster mass components, whose description is subject to a higher uncertainty compared to the determination of the mass profile of the galaxy-scale lens. As such, their inclusion may propagate systematics affecting the total cluster mass onto the determination of the galaxy-scale mass distribution.

To account for this uncertainty on the cluster mass distribution, we did not limit ourselves to considering the best-fit mass models, which may be affected by a systematic bias in the studied region. Instead, we extracted 100 random sets of parameter values from the Markov Chain Monte Carlo (MCMC) sampling of the posterior probability distribution of the reference cluster-scale SL model. We then fixed the mass distribution of the cluster-scale components and of all the other members to one of the 100 realisations thus obtained, and optimised the galaxy-scale lenses. We repeated the optimisation for each of the 100 mass models of the cluster. This bootstrapping procedure also allowed us to estimate the uncertainty on the determination of the parameters of the galaxy-scale lenses, and the degeneracies between them. Similarly to the reference models, we modelled the main lens galaxy of each galaxy-scale SL system as a spherical truncated dPIE. We also tested an elliptical non-truncated dPIE mass distribution to understand whether the preference for truncated models could arise from the insufficient azimuthal complexity of the spherical total mass model adopted. We performed all SL optimisations using \texttt{LensTool}. 

\subsection{MACS J0416.1$-$2403 member galaxy 8971}

The galaxy-scale SL system MACS J0416 ID14 was described in sub-section \ref{ss31}. Throughout the SL modelling procedure we adopted the multiple image catalogue presented in Table \ref{id14img}, comprising of eight multiple images from two background sources. All of the multiple images of the source ID14 are observed close to the cluster member 8971 (hereafter member 8971), at an average angular distance of $1.07''$: we thus focused on constraining its truncation radius. Cluster member 8980 significantly influences the multiple-image configuration, so we also optimised its mass distribution.

In \citetalias{bergamini22}, member 8971 was not included in the scaling relations adopted to describe the remaining cluster members, and it was modelled separately as a truncated dPIE profile. The values of the parameters of its total mass distribution are reported in Table \ref{id14b23}, with an uncertainty provided by the MCMC sampling of their marginalised posterior probability distribution. The ellipticity of the halo converges to the upper limit of its prior, indicating that it is poorly constrained and perhaps compensates for some unaccounted shear.

   \begin{table}
\caption{Best-fit values and $1\sigma$ errors of the parameters describing the cluster member 8971 in the SL model of MACS J0416 by \citetalias{bergamini22}.}             
\label{id14b23}      
\centering                          
\begin{tabular}{c c}        
\hline \noalign{\smallskip}                
Parameter & Value \\    
\noalign{\smallskip} \hline  \noalign{\smallskip}                      
           $\sigma \, (\mathrm{km \, s^{-1}})$ & $134^{+7}_{-6}$   \\
            \rule{0pt}{2.5ex}
            $r_t \, ('')$ & $18.6^{+8.5}_{-8.3}$ \\
            \rule{0pt}{2.5ex}
            $e$ & $0.52^{+0.06}_{-0.11}$     \\
            \rule{0pt}{2.5ex}
            $\theta_e \, (^{\circ})$ & $-40^{+20}_{-15}$    \\
 \noalign{\smallskip} \hline                                   
\end{tabular}
\end{table}

As anticipated, we first described member 8971 as a spherical truncated dPIE with vanishing core (hereafter SISt), whose centre is fixed at the light centre. The two free parameters are thus the velocity dispersion $\sigma$\footnote{included by \texttt{LensTool} via a fiducial velocity dispersion $\sigma_\mathrm{LT} = \sqrt{2/3} \sigma$.} and the truncation radius $r_t$. The alternative model, an elliptical non-truncated dPIE with a vanishing core (hereafter SIE), has three free parameters in $\sigma$, the ellipticity $e$\footnote{$e=(a^2-b^2)(a^2+b^2)^{-1}$, where $a$ and $b$ are the major and minor semi-axes of the ellipse.}, and the orientation angle $\theta_e$\footnote{Counter clockwise sexagesimal degree between the semi-major axis and the positive $x$ axis on the lens plane.}. As we wish to focus on studying the radial structure of member 8971, we considered simpler models for the mass distribution of the cluster member 8980. We tested spherical truncated and non-truncated isothermal models, and we noticed some degeneracy between the value of $r_t$ for member 8980 and the parameters describing member 8971. To avoid this, we chose to describe member 8980 with a SIS mass profile.

We optimised the two galaxy-scale SL models that we just outlined (with member 8971 parametrised as a SISt and a SIE respectively) for each of the 100 cluster-scale total mass distribution realisations obtained as described earlier. With this bootstrapping procedure, we obtained a set of 100 best-fit values for the parameters of the galaxy-scale SL system: we used the median, the 16th, and the 84th percentiles of the resulting distribution of best-fit values to obtain an estimate of their value and its uncertainty. Studying the marginalised distribution of the 100 best-fit parameters also provides us with insights on the degeneracy between them. Unlike in \citetalias{bergamini22}, where \texttt{LensTool} struggled to recover $r_t$ and the ellipticity of member 8971 at the same time, all the parameters of the SISt and SIE models are well constrained, with a low uncertainty. Their values are reported in Table \ref{id14g23}. In the case of the SISt mass profile, the total mass can be obtained as
\begin{equation}\label{massdpie}
    M = \frac{\pi \sigma^2 r_t}{G}.
\end{equation}
We can thus estimate a total mass value of $M=1.2^{+0.3}_{-0.1}\times10^{11} \, M_\odot$ for 8971.

   \begin{table}
\caption{Best-fit values and $1\sigma$ errors of the parameters of our galaxy-scale SL models of MACS J0416 ID14. In the first column 8971 is described as a SISt, in the second one as a SIE.}             
\label{id14g23}      
\centering                          
\begin{tabular}{c c c}        
\hline \noalign{\smallskip}                
Parameter & SISt & SIE \\    
\noalign{\smallskip} \hline  \noalign{\smallskip}                      
            Degrees of freedom & $8$ & $7$  \\
            \rule{0pt}{2.5ex}
           $\sigma_{8971} \, (\mathrm{km \, s^{-1}})$ & $164.9^{+6.8}_{-7.5}$ & $132.0^{+1.5}_{-1.2}$  \\
            \rule{0pt}{2.5ex}
            $r_{t,8971} \, ('')$ & $1.14^{+0.43}_{-0.20}$ & $-$\\
            \rule{0pt}{2.5ex}
            $e$ & $-$ & $0.20^{+0.03}_{-0.03}$     \\
            \rule{0pt}{2.5ex}
            $\theta_e \, (^{\circ})$ & $-$ & $57.5^{+3.8}_{-3.5}$    \\
             \rule{0pt}{2.5ex}
            $\sigma_{8980} \, (\mathrm{km \, s^{-1}})$ & $67.2^{+3.5}_{-5.8}$ & $45.6^{+3.1}_{-4.4}$    \\
            \rule{0pt}{2.5ex}
            $\langle \Delta_\mathrm{rms} \rangle \, ('')$ & $0.08$ & $0.04$ \\
 \noalign{\smallskip} \hline                                   
\end{tabular}
\end{table}

For each best-fit model, we estimate the value of $\Delta_\mathrm{rms}$, and we refer to the average value of $\Delta_\mathrm{rms}$ obtained for the 100 realisations of the model as $\langle \Delta_\mathrm{rms} \rangle$: we find $\langle \Delta_\mathrm{rms} \rangle=0.08''$ for the SISt models and $0.04''$ for the SIE model. We note that the SIE model allows for lower values of $\Delta_\mathrm{rms}$, which might be due to having one additional parameter compared to the SISt model. This may imply that a spherical mass distribution is too simple to describe the total mass distribution in the region, perhaps also due to unaccounted shear from the cluster-scale mass distribution. This is also suggested by the high value of ellipticity reported by \citetalias{bergamini22}.

\begin{figure}
   \centering
   \includegraphics[width=9.2cm]{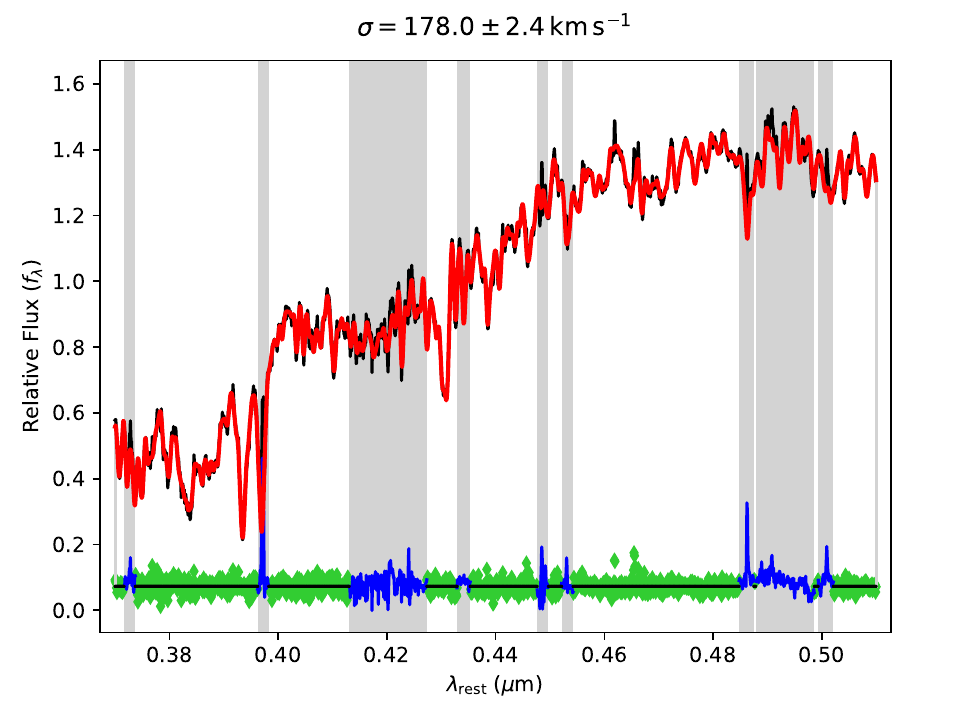}
      \caption{Fitting of the LOSVD of the MUSE spectrum of member 8971 with \texttt{pPXF}. The observed spectrum is shown in black, the red curve is the best-fit model, while the green points show the difference between the data and the model. The blue shaded regions along the wavelength axis were excluded in the fitting procedure, due to the presence of sky subtractions around emission lines or laser lines in the spectrum.}
         \label{vdispfitppxf}
\end{figure}

Given the significant difference in the best-fit value of $\sigma$ between the models, we compare them with the measured stellar line-of-sight velocity dispersion (LOSVD) for member 8971. The NE region of MACS J0416 was included in the MUSE Deep Lens Field \citet{vanzella21}, a very deep (17.1h integration time) MUSE observation with a seeing of approximately $0.6''$. The value of $\sigma$ recovered by SL models is a central three-dimensional mass density scale, while the value of LOSVD is a projected stellar velocity dispersion, which depends on the light distribution and on the point spread function of the observations, so the two values do not need to be exactly identical. We extracted the spectrum of member 8971 within a circular aperture centred in the centre of light of the galaxy and with a radius equal to the seeing of $0.6''$, to probe the central regions of the cluster galaxy. We measured the LOSVD using \texttt{pPXF} (penalised pixel-fitting) by \citet{cappellari04} and \citet{cappellari17,cappellari23}, comparing the observed spectra with a set of 463 UVB stellar templates from the X-shooter Spectral Library (XSL) DR2 \citep{gonneau20} with a signal-to-noise ratio ($S/N$) greater than $100 \, \mathrm{\r{A}^{-1}}$, combined and convolved with a LOSVD. The fit, whose results are shown in Fig. \ref{vdispfitppxf}, minimises a $\chi^2$ function between the observed spectrum and a model and provides a measured LOSVD value of $178.0 \pm 2.4 \, \mathrm{km \, s^{-1}}$, for an average $S/N$ value of $64.6$ over the spectrum. The very low error on the measured value only includes the statistical uncertainty of the final LOSVD fit and does not include possible systematics introduced by the choice of the stellar templates adopted to fit the spectrum. The velocity dispersion value found by our SISt galaxy-scale SL model is consistent within approximately $2\sigma$ with the measured LOSVD. The comparison with the value found by our SIE model is less straightforward due to the change in the surface area within a given iso-density contour at a fixed $\sigma$ determined by the introduction of the ellipticity through transformation defined in Eq. (\ref{ellvalue}). However, we note that our ellipticity value of $e=0.18$ implies a change in iso-density areas of only $1\%$ with respect to the spherical case, which allows us to compare the two values in first approximation. While the SL and the kinematic estimates of $\sigma$ do not probe exactly the same physical quantity, we would expect them to have similar values, and only our SISt model seems to allow for that. This comparison is not performed in the following sections for members 8785 and 3910, whose spectra have low $S/N$ values, probably due to a very faint lens magnitude and a lower MUSE exposure time, respectively. The determination of the $S/N$ threshold for a reliable LOSVD measurement requires further tests and larger samples, and will be presented in an upcoming work (Granata et al. in preparation). The predicted compactness of the cluster galaxy 8971 and the evidence for a truncated total mass density profile are discussed in Sect. \ref{s5}.

\subsection{MACS J0416.1$-$2403 member galaxy 8785}

The galaxy-scale SL system MACS J0416 ID16 was described in sub-section \ref{ss32}. Throughout the SL modelling procedure we adopted our new multiple image catalogue, presented in Table \ref{id16img} and comprising of six multiple images from two background sources. All six of the multiple images of the source ID16 are observed close to the cluster member 8785 (hereafter member 8785), at an average angular distance of only $0.43''$ from its centre. We thus focused on constraining the truncation radius of member 8785, and optimised the mass distribution of member 9129, which significantly influences the multiple-images configuration, as well.

In \citetalias{bergamini22}, member 8785 was modelled using the scaling relations (Eqs. \ref{scaling1} and \ref{scaling2}) leading to a velocity dispersion value $\sigma= 83.3^{+2.7}_{-6.7} \, \mathrm{km \, s^{-1}}$ and a truncation radius $r_t = 0.8^{+0.2}_{-0.1} \, ''$, resulting in a total mass $M=2.1^{+0.3}_{-0.2} \times 10^{10} \, M_\odot$. We first described member 8785 as a SISt, and we studied an alternative SIE model. Again, we tested several models for member 9129, and we chose a SIS mass profile to avoid degeneracies between parameters.

We ran the SISt and SIE models for each of the 100 cluster-scale total mass distributions extracted from the MCMCs. In Table \ref{id16g23}, we present the results of our bootstrapping procedure: the median values of the parameters of the best-fit galaxy-scale lensing models, and the 1$\sigma$ uncertainties derived from their 16th and 84th percentiles.

\begin{table}
\caption{Best-fit values and $1\sigma$ errors of the parameters of our galaxy-scale SL models of MACS J0416 ID16. In the first column 8785 is described as a SISt, in the second one as a SIE.}             
\label{id16g23}      
\centering                          
\begin{tabular}{c c c}        
\hline \noalign{\smallskip}                
Parameter & SISt & SIE \\    
\noalign{\smallskip} \hline  \noalign{\smallskip}                      
            Degrees of freedom & $5$ & $4$  \\
            \rule{0pt}{2.5ex}
           $\sigma_{8785} \, (\mathrm{km \, s^{-1}})$ & $57.6^{+0.3}_{-0.3}$ & $55.5^{+0.6}_{-1.2}$  \\
            \rule{0pt}{2.5ex}
            $r_{t,8785} \, ('')$ & $0.74^{+0.12}_{-0.08}$ & $-$\\
            \rule{0pt}{2.5ex}
            $e$ & $-$ & $0.18^{+0.08}_{-0.06}$     \\
            \rule{0pt}{2.5ex}
            $\theta_e \, (^{\circ})$ & $-$ & $110^{+10}_{-8}$    \\
             \rule{0pt}{2.5ex}
            $\sigma_{9129} \, (\mathrm{km \, s^{-1}})$ & $98.8^{+2.4}_{-2.2}$ & $88.3^{+1.6}_{-2.1}$    \\
            \rule{0pt}{2.5ex}
            $\langle \Delta_\mathrm{rms} \rangle \, ('')$ & $0.01$ & $0.01$ \\
 \noalign{\smallskip} \hline                                   
\end{tabular}
\end{table}

The offsets between the observed and model-predicted positions of the multiple images are extremely small, with a value of $\langle \Delta_\mathrm{rms} \rangle$, the average of $\Delta_\mathrm{rms}$ for the 100 model realisations, of $0.01''$ for both the SISt and the SIE models. This value is lower than the observational uncertainty on the determination of the positions of the multiple images. The distribution of the best-fit values of the free parameters, for the 100 cluster-scale mass distributions we consider, suggests that they are well constrained, with a low uncertainty and without clear degeneracies between them. Comparing our results for the SISt model with those from \citetalias{bergamini22}, we find lower values for $\sigma$ and $r_t$, corresponding to a lower total mass value of $M=1.0^{+0.2}_{-0.1} \times 10^{10} \, M_\odot$. This discrepancy may be due to a bias introduced by the scaling relations adopted in \citetalias{bergamini22}: cluster-scale SL models that describe the sub-halo component with the FP find that power-law scaling relations can over-predict the total mass \citepalias{granata22} and the velocity dispersion \citep{beauchesne23} of low- and intermediate-mass cluster galaxies. In Sect. \ref{s5}, we discuss the inferred compactness of member 8785.

\renewcommand{\thefootnote}{\fnsymbol{footnote}}

\subsection{MACS J1206.2$-$0847 member galaxy 3910}

The galaxy-scale SL system MACS J1206 ID14 was described in sub-section \ref{ss33}. Throughout the SL modelling procedure we adopted our new multiple image catalogue, presented in Table \ref{id141206img} and comprising of six multiple images from two background sources. All six of the multiple images of the source ID14 are observed close to the cluster member 3910 (hereafter member 3910), at an average angular distance of $1.11''$ from its centre: we thus focused on constraining its truncation radius. Cluster member 2541, at a distance of $6.86''$ from the centre of 3910, is the second brightest cluster galaxy, with a predicted total mass of $9.55 \times 10^{11} \, M_\odot$ in \citetalias{bergamini19}. Closer to 3910 is the low-mass ($M=1.97 \times 10^{9} \, M_\odot$) cluster member 3920. Due to its high total mass, we optimised the mass distribution of cluster member 2541 as well, whereas we kept the parameters describing 3920 fixed to the values predicted by cluster-scale modelling.

Member 3910 was described by \citetalias{bergamini19} with the two scaling relations (Eqs. \ref{scaling1} and \ref{scaling2}), leading to best-fit values of the velocity dispersion and of the truncation radius of $\sigma= 136.6^{+7.7}_{-6.6} \, \mathrm{km \, s^{-1}}$ and $r_t = 0.53^{+0.09}_{-0.07} \, ''$, for a total mass $M=4.2^{+0.4}_{-0.4} \times 10^{10} \, M_\odot$. Again, we first described 3910 as a SISt, and we tested an alternative SIE model. With regards to cluster member 2541, we note that in this case we are able to constrain its truncation radius without unforeseen parametric degeneracies, and thus we adopted a SISt model. 

We ran the SISt and SIE models for each of the 100 cluster-scale total mass distributions extracted from the MCMCs. In Table \ref{id141206g23}, we present the median values of the parameters of the best-fit galaxy-scale lensing models, as derived from the bootstrapping procedure, with 1$\sigma$ uncertainty derived from their 16th and 84th percentiles.

   \begin{table}
\caption{Best-fit values and $1\sigma$ errors of the parameters of our galaxy-scale SL models of MACS J1206 ID14. In the first column 3910 is described as a SISt, in the second one as a SIE.}             
\label{id141206g23}      
\centering                          
\begin{tabular}{c c c}        
\hline \noalign{\smallskip}                
Parameter & SISt & SIE \\    
\noalign{\smallskip} \hline  \noalign{\smallskip}                      
            Degrees of freedom & $3$ & $4$  \\
            \rule{0pt}{2.5ex}
           $\sigma_{3910} \, (\mathrm{km \, s^{-1}})$ & $129.2^{+5.1}_{-3.6}$ & $113.8^{+2.5}_{-3.0}$  \\
            \rule{0pt}{2.5ex}
            $r_{t,3910} \, ('')$ & $0.90^{+0.20}_{-0.22}$ & $-$\\
            \rule{0pt}{2.5ex}
            $e$ & $-$ & $0.77^{+0.06}_{-0.09}$     \\
            \rule{0pt}{2.5ex}
            $\theta_e \, (^{\circ})$ & $-$ & $96.53^{+0.29}_{-0.21}$    \\
             \rule{0pt}{2.5ex}
            $\sigma_{2540} \, (\mathrm{km \, s^{-1}})$ & $347^{+18}_{-16}$ & $399^{+1}_{-5}$\tablefootnote{The parameter value tends to the upper prior.}    \\
             \rule{0pt}{2.5ex}
            $r_{t,2540} \, ('')$ & $1.75^{+0.20}_{-0.18}$ & $0.49^{+0.03}_{-0.02}$    \\
            \rule{0pt}{2.5ex}
            $\langle \Delta_\mathrm{rms} \rangle \, ('')$ & $0.01$ & $0.03$ \\
 \noalign{\smallskip} \hline                                   
\end{tabular}
\end{table}
\renewcommand{\thefootnote}{\arabic{footnote}}

As in the case of MACS J0416 ID16, we find very small offsets between the observed and predicted multiple images, with an average $\Delta_\mathrm{rms}$ value of $\langle \Delta_\mathrm{rms} \rangle =0.01''$ for the SISt model and $0.03''$ for the SIE model. In spite of having fewer free parameters, the SISt model allows for a more accurate reconstruction of the lensing observables, and all parameters are estimated with a low statistical uncertainty, as clear from Table \ref{id141206g23}. On the other hand, the SIE model cannot constrain well the value of the velocity dispersion of member 2540, which tends to the upper limit of our prior, and the predicted ellipticity of member 3910 is very high\footnote{As such, the value of $\sigma_{3910}$ of the SISt model cannot be compared directly to that of the SIE model.}, suggesting that it might be needed to compensate for the lack of truncation in the total mass profile. Comparing our SISt model with \citetalias{bergamini19}, we find lower values for $\sigma$ and significantly higher values for $r_t$, for a total mass value of $M=6.3^{+1.0}_{-1.1} \times 10^{10} \, M_\odot$, higher than the best-fit value from \citetalias{bergamini19}. Again, the Faber-Jackson law adopted by \citetalias{bergamini19} may be biasing the value of $\sigma$ and therefore of the truncation radius. \citet{grillo14} use a one-parameter SIS mass profile to describe the cluster member, finding $\sigma=97\pm3 \, \mathrm{km \, s^{-1}}$. The lower value of $\sigma$ is expected, given the lack of a truncation of the mass profile. In the next section, we perform a more meaningful comparison, looking at the mass value within the effective radius.

\section{Analysis and discussion}\label{s5}

\subsection{Truncation radius of the cluster members}\label{ss51}
\begin{figure}
   \centering
   \includegraphics[width=9cm]{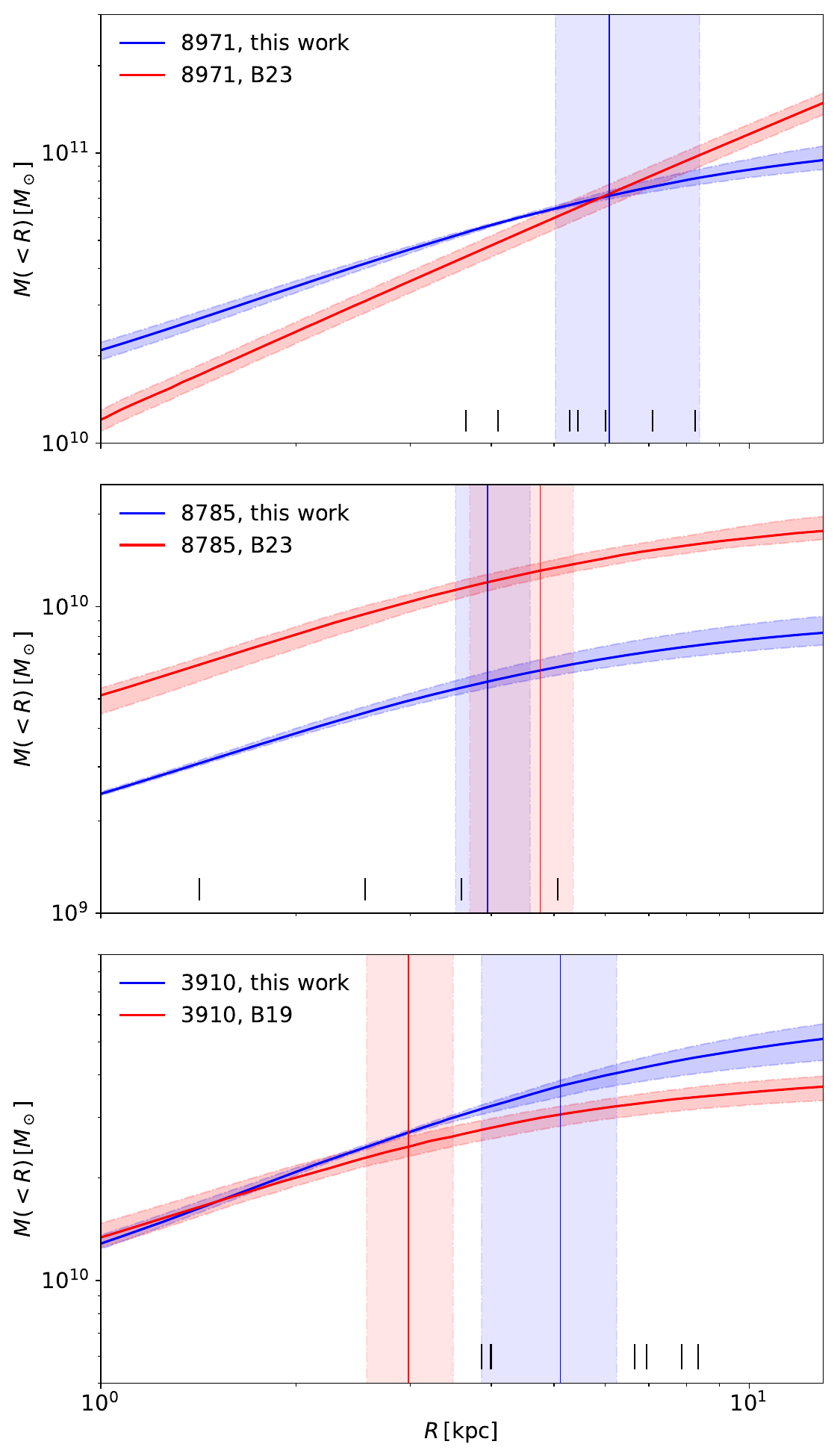}
      \caption{Projected cumulative total mass profiles for the lens galaxies studied in this work compared with the predictions of \citetalias{bergamini22} and \citetalias{bergamini19}. From the top to the bottom panel: member 8971, member 8785, and member 3910. The vertical bars indicate the inferred values of the truncation radius. Shaded regions indicate the 16th and 84th percentiles for the mass profile and the truncation radius. The projected distances of the observed multiple images from the lens centre are marked with vertical black lines.}
         \label{profiles}
\end{figure}

As detailed in the previous section, our galaxy-scale SL modelling procedure was aimed at estimating the truncation radius of three lens galaxies in massive clusters. The recovered values of $r_t$ for members 8971, 8785, and 3910 are $6.1^{+2.3}_{-1.1} \, \mathrm{kpc}$, $4.0^{+0.6}_{-0.4} \, \mathrm{kpc}$, and $5.2^{+1.3}_{-1.1} \, \mathrm{kpc}$, respectively. To better illustrate the radial scales at play, we built a cumulative total mass profile of the three members by comparing the 100 best-fit mass profiles obtained with the bootstrapping procedure and by taking the 50th percentile at each projected radius. We also estimated the statistical uncertainty on the total mass profile from the 16th and 84th percentiles. We performed the same procedure on the SL models of the three cluster members by \citetalias{bergamini22} and \citetalias{bergamini19}, using their MCMC sampling of the parametric posterior probability distribution to quantify the uncertainty on their total mass distributions. These profiles are shown in Fig. \ref{profiles}. We note that this procedure only accounts for the statistical uncertainty on the mass distribution parameters, and not for the possible systematics.

\begin{figure*}
   \centering
   \includegraphics[width=18.5cm]{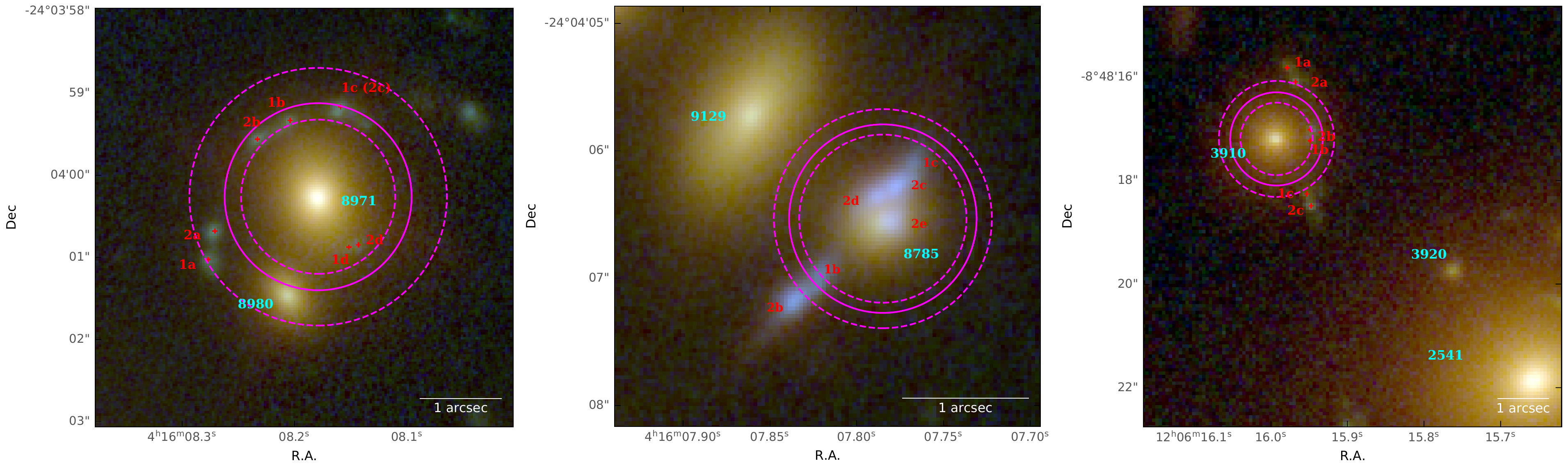}
      \caption{Truncation radius of members 8971, 8785, and 3910 superimposed to the RGB cutout of the respective galaxy-scale SL system. We show the best fit value as a solid line and the $1\sigma$ uncertainty range with dashed lines.}
         \label{f814hst}
   \end{figure*}

As shown in Figs. \ref{profiles} and \ref{f814hst}, the measured values of the truncation radius are, for all three members, within the range of the observed positions of the multiple images. In this radial range, SL allows for the highest accuracy in the reconstruction of the total mass profile of a lens, and is therefore more sensitive to its slope. In all three cases, we have tested an alternative non-truncated SIE mass parametrisation to inquire whether the truncated profile could actually be compensating for an insufficient description of the lens azimuthal structure. Despite having $4$ free parameters, as opposed to the $3$ of the SISt model, only in one of the three cases (member 8971), the SIE model leads to a slightly lower value of $\Delta_\mathrm{rms}$. In one case (member 3910), the SIE model predicts a very high value of the lens ellipticity, which is not suggested by the light distribution. In conclusion, in all three systems we find that a spherical truncated total mass distribution for the lens galaxy is able to reproduce the SL observations with a small $\Delta_\mathrm{rms}$ and provide the value of $r_t$ with a low statistical error. In two out of the three cases, a non-truncated model with a higher number of parameters does not improve the description of the system, in one case leading to unrealistic parameter values.

\begin{figure}
   \centering
   \includegraphics[width=9cm]{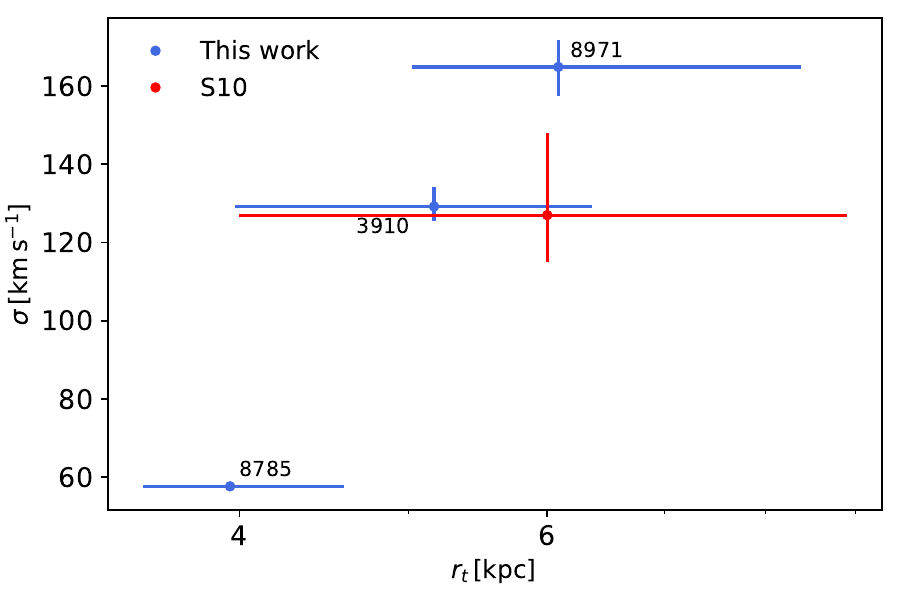}
      \caption{Relation between the values of the velocity dispersion and of the truncation radius for the three cluster members included in this study (in blue) and for a satellite of SL2S J08544$-$0121, as measured by \citet{suyu10} (in red). The inferred uncertainty on the velocity dispersion of member 8785 is too small to be visible in this plot.}
         \label{suyu10}
\end{figure}

\citet{suyu10} performed a similar truncation radius measurement for a satellite galaxy ($z=0.351$) of the massive elliptical lens SL2S J08544$-$0121, which influences the shape of an Einstein ring determined by the main lens. They find a value of $r_t$ close to its theoretically predicted tidal radius. Figure \ref{suyu10} shows that the value of $r_t$ estimated in \citet{suyu10} seems to agree with the $\sigma$-to-$r_t$ relation found in our work. \citet{monna15} repeated the procedure for the CLASH-VLT galaxy cluster Abell 383. They combined SL with observational priors on the observed velocity dispersion for two high mass (total mass greater than $10^{12} \, M_\odot$) members close to two lensed arcs of the same sources. The recovered value of $r_t$ is greater than $50 \, \mathrm{kpc}$ for both galaxies, significantly higher than predicted by the scaling law calibrated for the other cluster members of Abell 383. The authors suggest that very bright members may have not undergone strong stripping as a result of being a central galaxies prior to accretion on the cluster. Both \citet{suyu10} and \citet{monna15} based their SL modelling procedure on surface brightness reconstruction of extended arcs, finding that this significantly improves the accuracy of the constraints on the value of the truncation radius. In this work, we only used the centroid position of the multiple images, rather than their full surface brightness. However, unlike in \citet{suyu10} and \citet{monna15}, the cluster-scale mass distribution determines several different multiple images of the background source close to the galaxy-scale lens, rather than a single extended lensed arc. In our case, several multiple images are observed at different projected distances from the lens centre, providing us with detailed information about the galaxy total mass profile and its slope.

Finally, we tested whether the inferred values of $r_t$ may be biased by the parametrisation chosen by \citetalias{bergamini22} and \citetalias{bergamini19} for the total mass profile of the haloes included in the model. To do so, we built an alternative cluster-scale model for MACS J1206 in which all of the cluster members are described with singular isothermal sphere (SIS) mass profiles (i.e. with infinite $r_t$) and repeated the statistical analysis described above. We find that this does not influence significantly the estimated parameter values for the lens galaxy 3910, suggesting that our modelling procedure is robust with respect to the parametrisation choices adopted for the remaining cluster mass components.

\subsection{Compactness of the cluster galaxies}

Galaxy-scale SL events in massive clusters allowed us to infer the physical properties of some selected cluster galaxies without relying on the scaling laws typically adopted to describe them. Cluster-scale SL modelling is mostly sensitive to the total mass of the cluster galaxies, rather than to the details of their mass density profiles. As such, $\sigma$ and $r_t$ suffer from a strong degeneracy and cannot be separately constrained in absence of an observational prior: this significantly limits the insights on the compactness of the cluster galaxies.   
To break the degeneracy between $\sigma$ and $r_t$, \citetalias{bergamini19} and \citetalias{bergamini22} obtained a kinematic prior on the value of the slope of the Faber-Jackson law (marked as $\alpha$ in Eq. \ref{scaling1}). Assuming a total mass-to-light ratio $M/L \propto L^\gamma$ leads to $\beta=\gamma-2\alpha+1$ ($\beta$ is defined in Eq. \ref{scaling2}). From Eq. (\ref{massdpie}), one can then derive a total mass-to-$\sigma$ scaling law
\begin{equation} \label{msigscaling}
    M \propto \sigma^{\frac{1+\gamma}{\alpha}}.
\end{equation}
\citetalias{bergamini19} and \citetalias{bergamini22} assumed $\gamma=0.2$, as suggested by the FP. The reference values of the scaling laws ($\sigma^\mathrm{ref}$ and $r^\mathrm{ref}_t$ in Eqs. \ref{scaling1} and \ref{scaling2}) are mostly determined by the high-mass cluster members, which have a stronger influence on the positions of the observed multiple images. On the other hand, once $\gamma$ is fixed, the value of $\alpha$ fixes the slopes of the two laws, determining the description of the total mass distribution of the low- and intermediate-mass cluster members. 

\begin{figure}
   \centering
   \includegraphics[width=9cm]{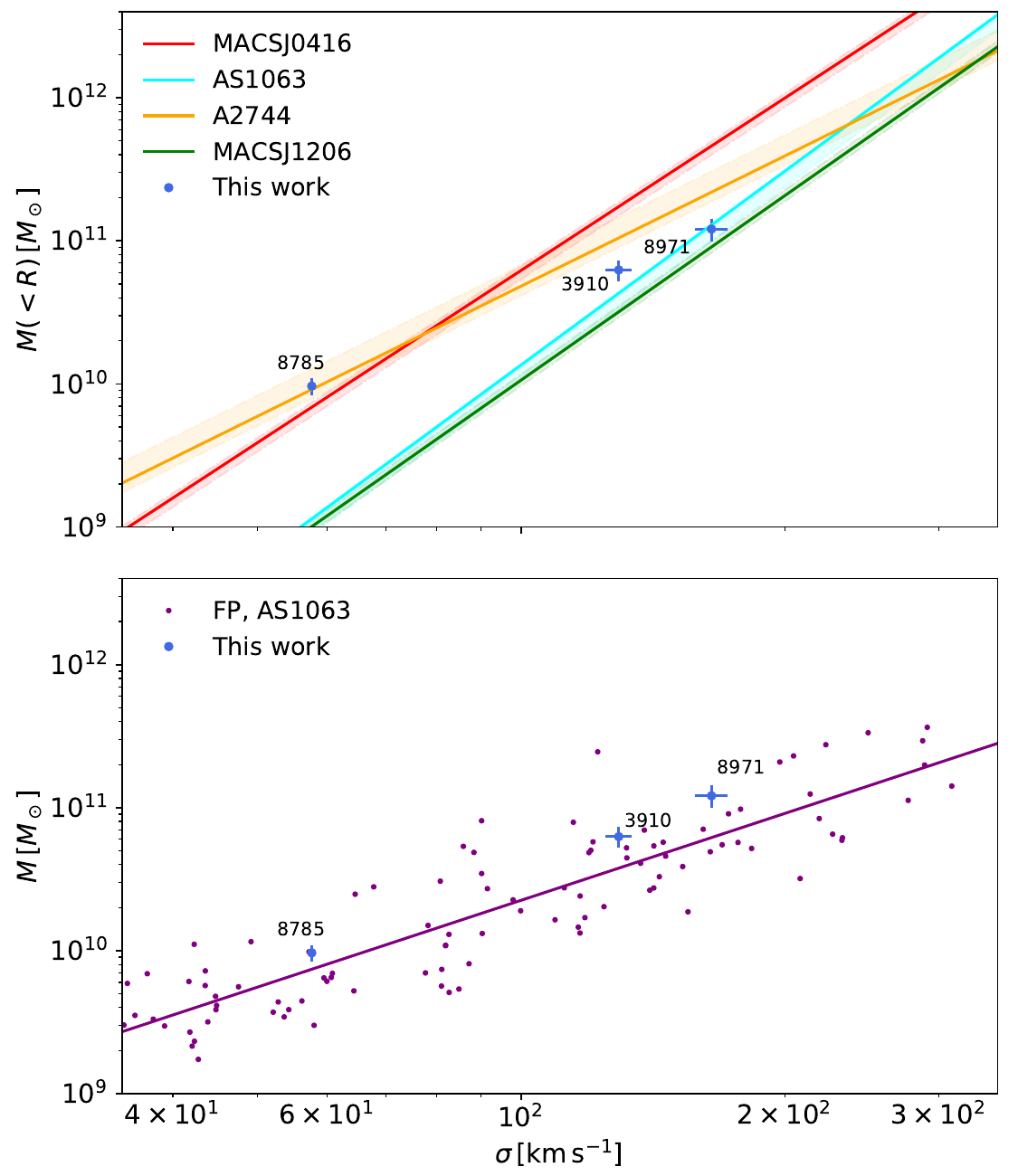}
      \caption{Comparison between the $M$-to-$\sigma$ relation estimated by our analysis and those adopted for the cluster galaxies in the SL models of MACS J0416 \citepalias{bergamini22}, AS1063 \citepalias{bergamini19,granata22}, MACS J1206 \citepalias{bergamini19}, and A2744 \citep{bergamini23}. Top panel: comparison with the SL models describing the cluster galaxies according to Eqs. (\ref{scaling1}) and (\ref{scaling2}). Bottom panel: comparison with the SL model of AS1063 by \citetalias{granata22}, based on the FP.}
         \label{scaling}
\end{figure}

Considering the reference sample of clusters included in \citet{meneghetti20}, \citetalias{bergamini19} find $\alpha=0.28^{+0.02}_{-0.02}$ for MACS J1206 and $\alpha=0.27^{+0.04}_{-0.04}$ for AS1063, while \citet{bergamini21} and \citetalias{bergamini22} find $\alpha=0.30^{+0.03}_{-0.03}$ for MACS J0416, corresponding to $M\propto\sigma^{4.3}$, $M\propto\sigma^{4.4}$, and $M\propto\sigma^{4.0}$, respectively. The same procedure was adopted by \citet{bergamini23} for the HFF cluster Abell 2744 (A2744), finding a value of $\alpha=0.40^{+0.03}_{-0.03}$ (implying $M\propto\sigma^{3.0}$). In the top panel of Fig. \ref{scaling}, we compare the $M$-to-$\sigma$ scaling law for these four cluster scale SL models with the values of total mass and velocity dispersion for the galaxy-scale SL systems. In cluster-scale models, only the values $\sigma^\mathrm{ref}$ and $r^\mathrm{ref}_t$ are optimised: we estimated the uncertainty on the determination of the $M$-to-$\sigma$ relation based on the MCMC sampling of the posterior probability distribution for these two parameters. We notice that the scaling relations used in cluster-scale models have significantly different normalisation and slope values. In particular, the SL model of MACS J0416 consistently predicts higher mass values at a fixed $\sigma$ compared to those of MACS J1206 and AS1063. On the other hand, the model of A2744 has a significantly lower slope, predicting higher total mass values at low $\sigma$. Looking at the three cluster members studied in this work, only the $M$-to-$\sigma$ relation found for A2744 seems to represent well their compactness.

In the SL model of AS1063 by \citetalias{granata22}, the values of $\sigma$ were fixed from the observed luminosity and half-light radius $R_e$ of the cluster galaxies, while a proportionality law was calibrated between the observed $R_e$ and $r_t$. As such, the $M$-to-$\sigma$ relation is not a power-law and is able to include a realistic scatter. As shown by the bottom panel of Fig. \ref{scaling}, the relation significantly differs from those obtained with a power-law approach: a bi-logarithmic fit of the relation predicts a slope of $2.0$. The panel also shows that the three galaxy-scale lenses that we have modelled in this work lie within the scatter of the scaling relation derived by \citetalias{granata22} using the FP. While a sample of three objects is very small, it is interesting to note that the $M$-to-$\sigma$ relation for the galaxy scale lenses, derived exclusively with SL, is very close to the predictions of \citetalias{granata22}, where lensing observables are only used to estimate the ratio between $R_e$ and $r_t$. A more complex description of the cluster galaxies based on the FP leads to inferred properties which are compatible with our analysis. \citet{beauchesne23} recently modelled AS1063 adopting an intermediate approach between \citetalias{bergamini19} and \citetalias{granata22}, where galaxies are described with the Faber Jackson law or the FP depending on the observations available for them. The value of $\alpha$ was optimised together with the parameters of the FP to avoid inconsistent slopes. Choosing $\gamma=0$, they find $\alpha=0.34$, obtaining to $M\propto\sigma^{3}$. On the other hand, fixed power-law scaling relations do not seem to be able to correctly describe the spatial structure of the cluster galaxies on the whole mass range included in SL models. As shown by Fig. \ref{profiles}, we predict significantly different properties for members 8785 and 3910 compared to \citetalias{bergamini22} and \citetalias{bergamini19}: the former has a similar value of $r_t$ but a higher $\sigma$, while the latter has a significantly larger $r_t$. 

\begin{figure}
   \centering
   \includegraphics[width=9cm]{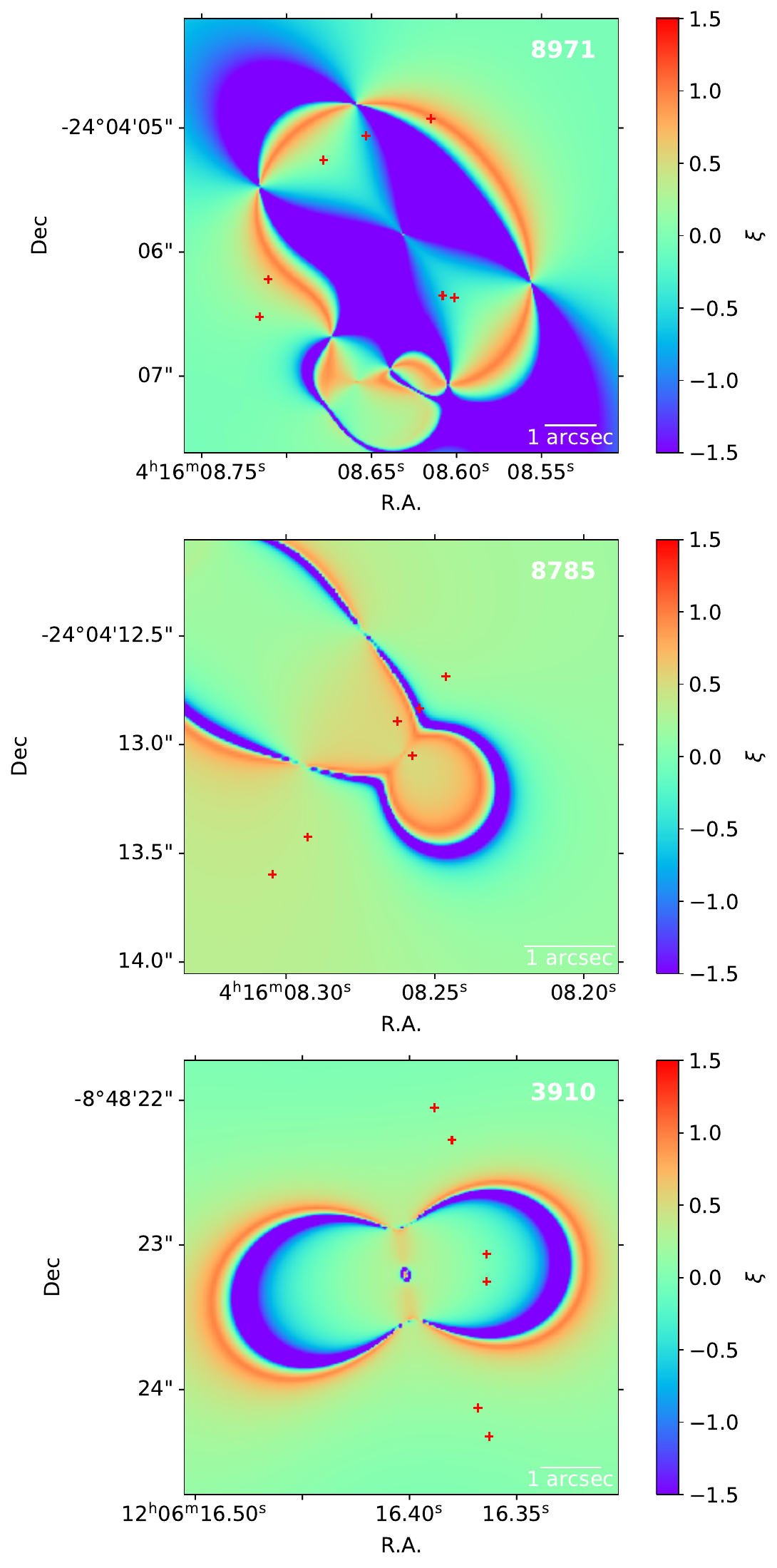}
      \caption{Comparison between the magnification values close to the three galaxy-scale lenses predicted in this work and in \citetalias{bergamini22} and \citetalias{bergamini19}. The colour map is based on the value of $\xi=\log \left( \frac{\mu - \mu_\mathrm{CS}}{\sigma_\mu} \right)$ close to members 8971, 8785, and 3910, respectively. The red crosses indicate the observed positions of the multiple images included in this work.}
         \label{magnification}
\end{figure}

These differences may have a non-negligible impact on the magnification factor predicted close to the galaxy-scale lenses, which connects the observed and the unlensed magnitude for the multiply imaged sources. We measured the magnification predicted by the three galaxy-scale models. For each of them, we used the 100 models from our boot-strapping procedure to estimate the uncertainty on the magnitude value. In Fig. \ref{magnification}, we map the value of $\xi=\log \left( \frac{\mu - \mu_\mathrm{CS}}{\sigma_\mu} \right)$, where $\mu$ is the magnification predicted by our models, $\sigma_\mu$ is the uncertainty on its value, and $\mu_\mathrm{CS}$ is the magnification from the cluster-scale models (by \citetalias{bergamini22} and \citetalias{bergamini19}). The upper panel of Fig. \ref{magnification} shows that our work and \citetalias{bergamini22} predict significantly different magnifications close to member 8971. 
However, these differences are probably due to the different mass parametrisation chosen by the two models and are less pronounced close to the positions of the multiple images, where the total mass distribution is better constrained. As expected, close to members 8785 and 3910, the value of $\xi$ is closer to zero as a consequence of choosing the same mass parametrisation in the two models. In all three cases, with few exceptions, the difference between the predicted values of the magnification is relatively small close to the multiple images, where the total mass distribution of the lens is better reconstructed. In conclusion, the systematics related to the mass modelling of the cluster members affect the predicted magnification map of lens clusters close to member galaxies. These predictions are more robust in proximity of the observed positions of the multiple images, but a more realistic description of the total mass properties of lens galaxies can benefit the accuracy of the studies of high-redshift lensed sources.

\subsection{Comparison with cosmological simulations}

\begin{figure}
   \centering   \includegraphics[width=9cm]{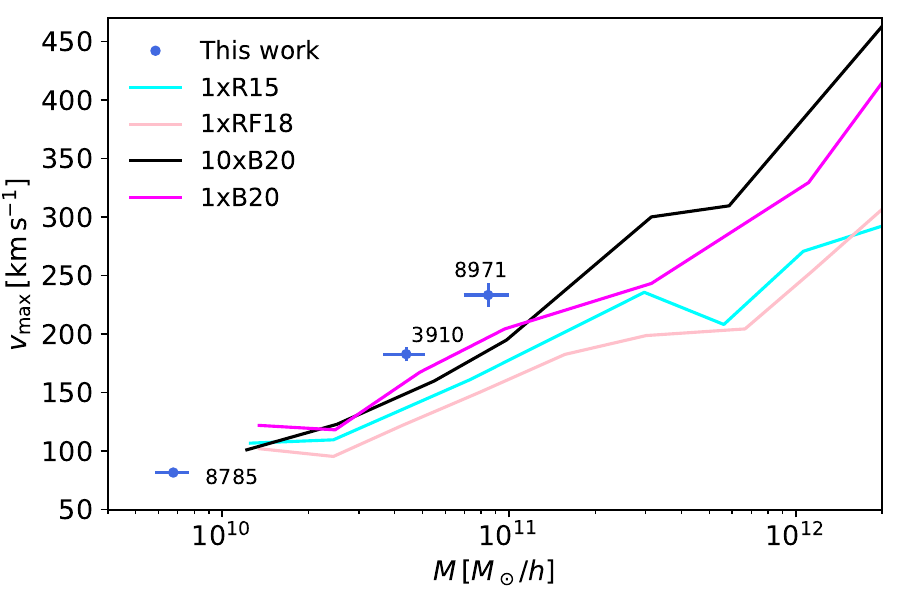}
      \caption{Comparison between the $v_\mathrm{max}$-to-$M$ relation obtained from our analysis and those predicted by the cosmological hydrodynamical simulations described in \citet{ragagnin22}. The naming convention for the different suites is presented in the text.}
         \label{vmax}
\end{figure}

In this sub-section, we compare the compactness of the cluster members as obtained from our galaxy-scale SL models with the theoretical predictions of cosmological simulations. This study offers an excellent opportunity to contrast the properties of observed and simulated galaxy-scale lenses in massive clusters (although we note that member 8785 is not the main lens responsible for the secondary critical line that produces system ID16). As in \citet{meneghetti20}, we adopt the maximum circular velocity of the cluster members, defined as 
\begin{equation}
    v_\mathrm{max}= \mathrm{max}\left(\sqrt{\frac{GM(<r)}{r}}\right),
\end{equation} 
where $v_\mathrm{max} = \sqrt{2} \sigma$ for an isothermal model, as a proxy for the compactness of cluster galaxies \citep[see][]{meneghetti20,meneghetti22,meneghetti23,ragagnin22}: more compact objects have higher values of $v_\mathrm{max}$ at a fixed total mass. In Fig. \ref{vmax}, we compare the $v_\mathrm{max}$-to-$M$ relation for our work with those found in \citet{ragagnin22} for a set of zoom-in re-simulations of the Dianoga suite \citep{planelles14,rasia15} of simulated galaxy clusters. These setups differ from one another in terms of their softening and feedback schemes. As in \citet{ragagnin22}, we refer to the three models considered as R15 \citep[presented in][]{rasia15}, RF18 \citep[presented in][]{ragone18}, and B20 \citep[presented in][]{bassini20}. The setups are also referred to as 1x or 10x if they have the same mass resolution as the Dianoga suite, or a ten times lower particle mass, respectively.

Figure \ref{vmax} shows that members 8971 and 3910 both have a maximum circular velocity value higher than those for simulated sub-haloes with the same total mass, irrespective of the feedback scheme or the resolution considered in \citet{ragagnin22}, indicating a level of compactness higher than that predicted by simulations. This seems to suggest that the discrepancy between SL models and simulations cannot be entirely ascribed to systematics affecting the former, as a result of the adoption of power-law scaling relations to describe the cluster members, as noted already by \citetalias{granata22}. Member 8785 falls in a total mass range ($M<10^{10} \, M_\odot$) which was excluded from the analyses of \citet{meneghetti20}, because the current cosmological simulations do not have enough mass resolution.

\renewcommand{\thefootnote}{\arabic{footnote}}

\subsection{Stellar mass of the cluster members}

In this sub-section, we study the stellar-to-total mass fraction of the cluster members. Measuring its value within the effective radius is an important probe of the interplay between the effects of the baryonic feedback processes and the gravitational potential of DM during galaxy formation \citep[see][]{shajib22,smith20}, and of stellar populations in early-type galaxies. These processes are particularly significant for lower-mass galaxies, which should have a higher stellar-to-total mass fraction compared to the very massive early-type galaxies that dominate the samples of lens galaxies. Our work significantly extends the typical mass range of current studies, similarly to what will be performed with the upcoming samples of galaxy-scale lenses unveiled by \textit{Euclid} and LSST.

   \begin{figure*}
   \centering
\includegraphics[scale=0.58] {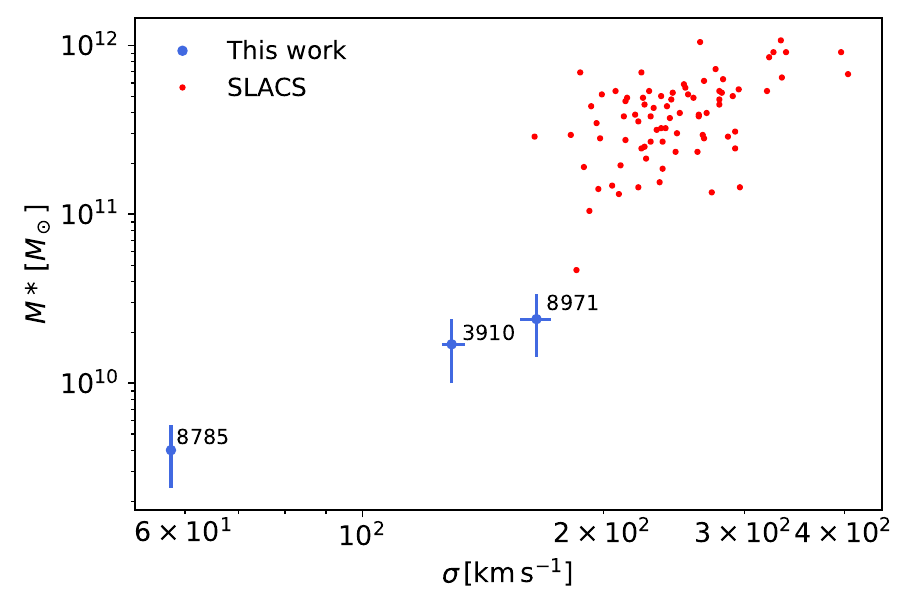} 
\includegraphics[scale=0.58] {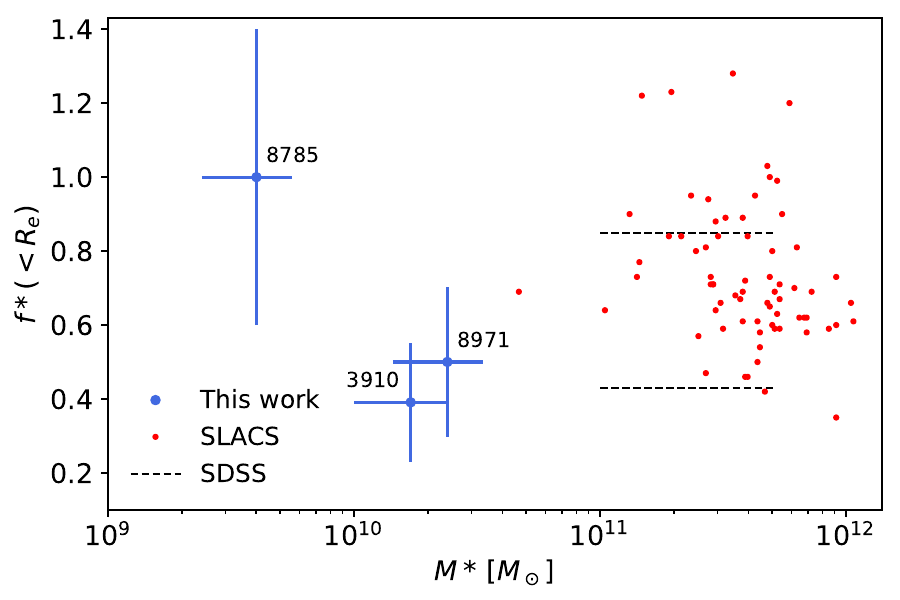} 
	
   \caption{Stellar mass of lens galaxies: comparison between this work and the 85 SLACS lens galaxies presented in \citet{auger09}. Left panel: Stellar mass as a function of the velocity dispersion. Right panel: Stellar-to-total mass fraction measured within the effective radius; the 68\% confidence interval found by \citet{grillo10} for a selected sample of $2\times10^5$ SDSS early-type galaxies is also shown.}
   \label{smass}
   \end{figure*}

The values of the total stellar mass of the three lens galaxies have been measured in recent works. For members 8971 and 8785, we followed the best-fit relation between the stellar mass and the magnitude in the HST F160W band found by \citet{grillo15}, $\log \left(M^* (M_\odot) \right) = 18.541 - 0.416 \times F160W$, where $M^*$ is the stellar mass of the cluster member. To consider the scatter about this mean relation, we chose a $40\%$ uncertainty on the derived stellar mass values of $(2.4\pm1.0)\times10^{10} \, M_\odot$, and $(4.0\pm1.6)\times10^9 \, M_\odot$, respectively. For member 3910, we adopted the value measured by \citet{grillo14} of $(1.7\pm1.0)\times10^{10} \, M_\odot$. In both cases, the stellar mass measurements were based on the HST spectral energy distribution (SED) of the lens galaxies, using composite stellar population models, and a Salpeter \citep[][]{salpeter55} stellar initial mass function (IMF). In Fig. \ref{smass}, we show the relation between the stellar mass and the velocity dispersion for the three cluster members studied in this work and for the 85 SLACS lenses presented in \citet{auger09}, which are representative of the currently known population of lens galaxies. We note that the SLACS lenses were modelled with non-truncated total mass profiles, which could affect the recovered values of the velocity dispersion. The figure showcases the significant extension of the range of stellar mass and velocity dispersion values probed in this work compared to current samples of lens galaxies.

Using \texttt{GALFIT} on the HFF (for MACS J0416) or CLASH (for MACS J1206) F814W band images, we measured the effective radii, $R_e$, of the three galaxies, finding $1.41\pm0.02 \, \mathrm{kpc}$\footnote{Compatible with the value measured by \citet{tortorelli23}.}, $0.77\pm0.03 \, \mathrm{kpc}$, and $2.13\pm0.09 \, \mathrm{kpc}$, respectively. They correspond to a ratio between the truncation and the half-light radius $\frac{r_t}{R_e}$ of $4.3 \pm 1.6$, $5.13\pm0.72$, and $2.77\pm0.47$, all higher than the average value of $2.3$ found by \citetalias{granata22} for the cluster members of AS1063, although they fall within the scatter around their mass-to-$\sigma$ relation. From the total mass profile derived for each of the 100 best-fit models of the three galaxies, we obtained the total mass enclosed within the effective radius $M(<R_e)$ and its uncertainty. The stellar-to-total mass fraction within the effective radius for the three members is therefore
\begin{equation}
    f^*(<R_e)=\frac{M^*/2}{M(<R_e)}.
\end{equation}
We find $0.51\pm0.21$, $1.0\pm0.4$, and $0.39\pm0.16$, for members 8971, 8785, and 3910, respectively.

We compare our values of the stellar-to-total mass fraction as a function of the stellar mass with the analogous relation found for the 85 SLACS lens galaxies presented in \citet{auger09} \citep[similar results had previously been obtained by][]{grillo08}. We take their stellar mass values measured with a Salpeter stellar IMF and the stellar-to-total mass fraction within the effective radius measured in the rest-frame $V$-band (comparable with the F814W band for the two clusters considered in this work). As clear from Fig. \ref{smass}, we probe a lower stellar mass range, but find compatible values between the two samples. Our values agree with the results of \citetalias{granata22} for the cluster members of AS1063. We also included in Fig. \ref{smass} the values found, starting from the same hypotheses, by \citet{grillo10} for a selected sample of $2\times10^5$ SDSS early-type galaxies, which we find to be compatible both with our work and with that by \citet{auger09}. These results suggest that the tidal truncation to which the three lens galaxies are subject to, by virtue of the dense cluster environment in which they reside, only marginally affects their structure within the effective radius. This is in agreement with the conclusions of \citet{grillo10b}, \citet{grillo14}, \citet{parry16}, and \citetalias{granata22}, who compared the stellar fraction within the effective radius of cluster and field early-type galaxies.

\section{Conclusions}\label{s6}

In this article we have presented the measurement with SL of the truncation radius of three cluster galaxies. We considered the reference sample of galaxy clusters included in the analysis by \citet{meneghetti20} and selected galaxy-scale SL systems with a clear morphology and several multiple images close to one or a few member galaxies. We chose to focus on members 8971 and 8785 of MACS J0416, and member 3910 of MACS J1206. We built galaxy-scale SL models for the three cluster members and for the neighbouring galaxies which influence the lensing system the most. 

We accounted for the lensing effects of the remaining mass components of the cluster according to the predictions of the most recent and accurate SL models of MACS J0416 and MACS J1206, presented in \citetalias{bergamini22} and \citetalias{bergamini19}, respectively. To properly consider the uncertainty on the total mass distribution of the two clusters, we sampled the posterior probability distribution of the parameters of the two models, and extracted 100 points, corresponding to 100 realisations of the cluster-scale mass distribution. For each of them, we optimised the models of the mass distribution of the three galaxy-scale lenses. This bootstrapping procedure allowed us to obtain a realistic estimate of the uncertainty on the lens parameters and of the degeneracy between them. We described the three members on which we focus our analysis with spherical truncated profiles and test alternative ellipsoidal non-truncated models. The main conclusions of our analyses are summarised as follows:

\begin{enumerate}
    \item We measured a truncation radius value of $6.1^{+2.3}_{-1.1} \, \mathrm{kpc}$, $4.0^{+0.6}_{-0.4} \, \mathrm{kpc}$, and $5.2^{+1.3}_{-1.1} \, \mathrm{kpc}$ for members 8971, 8785, and 3910, respectively. These values correspond to a total mass of $M=1.2^{+0.3}_{-0.1}\times10^{11} \, M_\odot$, $M=1.0^{+0.2}_{-0.1} \times 10^{10} \, M_\odot$, and $M=6.3^{+1.0}_{-1.1} \times 10^{10} \, M_\odot$, respectively, and to velocity dispersion values of $164.9^{+6.8}_{-7.5} \, \mathrm{km \, s^{-1}}$, $57.6^{+0.3}_{-0.3} \, \mathrm{km \, s^{-1}}$, and $129.2^{+5.1}_{-3.6} \, \mathrm{km \, s^{-1}}$, respectively.
    \item The values of $r_t$ are well constrained, with a low statistical uncertainty. We compare our results with those of \citet{suyu10} for the satellite galaxy of the lensing system SL2S J08544$-$0121, finding very similar $r_t$ values for galaxies in the same total mass range.
    \item In the case of member 8971, the SIE model leads to a lower value of $\Delta_\mathrm{rms}$, but the comparison between the SL-derived $\sigma$ and the measured LOSVD value of $178.0 \pm 2.4 \, \mathrm{km \, s^{-1}}$ strongly favours the SISt model.
    \item In the other two instances, SIE models do not lead to an improved accuracy of the description of the SL observations, in spite of their more complex azimuthal structure and a higher number of free parameters. In the case of member 3910, the parameters of the non-truncated models are not well constrained and show clear degeneracies.
    \item Our inferred values of $\sigma$ and $r_t$ for the three cluster galaxies differ significantly from the results of \citetalias{bergamini22} and \citetalias{bergamini19}, especially in the case of members 8785 and 3910, where they were derived with power-law scaling relations with respect to the galaxy total luminosity.
    \item We compare our results with the total-mass-to-$\sigma$ relations for MACS J0416, MACS J1206, AS1063, and A2744 from \citetalias{bergamini22}, \citetalias{bergamini19}, and \citet{bergamini23}, obtained with the power-law approach. We find that the scaling relations cannot consistently describe all three members studied in this work. Our results instead agree with the  mass-to-$\sigma$ relation derived by \citetalias{granata22} for AS1063, based on the FP relation and showing a larger scatter.
    \item We juxtapose the estimated compactness of the three lens galaxies with the predictions of the hydrodynamical cosmological simulation suites presented in \citet{ragagnin22}, which differ in feedback and softening scheme, and mass  resolution. Members 8971 and 3910 fall in the total mass range included in the analyses performed on the simulations. Their measured compactness is higher than what is found for simulated sub-haloes of the same total mass, independently of the simulation set-up considered, confirming the discrepancy between observations and simulations first reported in \citet{meneghetti20}.
    \item We measured the stellar mass and the effective radius, and the stellar-to-total mass fraction within the effective radius for the three cluster galaxies. For the latter parameter, we find $0.51\pm0.21$, $1.0\pm0.4$, and $0.39\pm0.16$ for members 8971, 8785, and 3910, respectively. Our values span the same range as those observed for the members of AS1063 by \citetalias{granata22}, for the 85 SLACS lens galaxies presented in \citet{auger09}, and for a selected sample of early-type SDSS galaxies by \citet{grillo10}, suggesting that the tidal truncation of cluster galaxies does not significantly affect their structure within the effective radius.
\end{enumerate}
As clear from Fig. \ref{smass}, our work significantly extends the mass range probed by current SL studies of early-type galaxies, towards the regimes that will be systematically explored by the upcoming lens surveys with the \textit{Euclid} and the \textit{Rubin} telescopes. On-going integral-field observations with the \textit{James Webb} Space Telescope Near Infrared Spectrograph will provide us with spatially resolved kinematic constraints of galaxy-scale lenses in clusters, allowing for a more detailed reconstruction of the mass structure of member galaxies.


\begin{acknowledgements}
We thank the anonymous referee for some useful suggestions that helped us improve the paper. We  acknowledge  financial  support  by  PRIN-MIUR 2017WSCC32 (P.I:: P. Rosati),  PRIN-MIUR 2020SKSTHZ  (P.I.: C. Grillo),  INAF  main-stream  1.05.01.86.20 (P.I.:  M.  Nonino)  and  INAF  1.05.01.86.31 (P.I.: E. Vanzella). MM was supported by INAF Grant The Big-Data era of cluster lensing. MM also acknowledges support from the Aspen Center for Physics and Simons Foundation. This work uses the following software packages:
\href{https://projets.lam.fr/projects/lenstool/wiki}{\texttt{Lenstool}}
\citep{jullo07, jullo09},
\href{https://github.com/astropy/astropy}{\texttt{Astropy}}
\citep{astropy1, astropy2},
\href{https://github.com/matplotlib/matplotlib}{\texttt{matplotlib}}
\citep{matplotlib},
\href{https://github.com/numpy/numpy}{\texttt{NumPy}}
\citep{numpy1, numpy2},
\href{https://pypi.org/project/ppxf/}{\texttt{pPXF}}
\citep{cappellari04,cappellari23},
\href{https://www.python.org/}{\texttt{Python}}
\citep{python},
\href{https://github.com/scipy/scipy}{\texttt{Scipy}}
\citep{scipy}.

\end{acknowledgements}

%
\bibliographystyle{aa} 
\bibliography{bibl.bib} 
%

\end{document}